\documentclass[sigconf]{acmart}

\usepackage{balance}    %
\usepackage{color}
\usepackage{color, colortbl}
\usepackage{array}
  
\usepackage{caption}
\usepackage{subcaption}

\usepackage{multirow}
\usepackage{siunitx}

\usepackage{xspace}

\usepackage{graphicx}
\usepackage{xcolor}
\usepackage{mwe}
\usepackage{blindtext}
\usepackage{todonotes}
\usepackage{draftfigure}

\definecolor{Gray}{gray}{0.97}
\definecolor{MedGray}{gray}{0.9}
\definecolor{greytext}{gray}{0.5}
\definecolor{DarkGreen}{rgb}{0.0, 0.5, 0.0}
\definecolor{PFGreen}{rgb}{0.0, 0.5, 0.0}
\definecolor{lightGreen}{rgb}{0.8, 0.9, 0.8}
\definecolor{CadmiumGreen}{rgb}{0.0, 0.42, 0.24}
\definecolor{DarkKhaki}{rgb}{0.74, 0.72, 0.42}
\definecolor{DarkRed}{rgb}{0.7, 0.2, 0.2}
\definecolor{Purple}{rgb}{0.7,0.0,0.7}
\definecolor{Brown}{rgb}{0.7,0.3,0}
\definecolor{Orange}{rgb}{1, 0.5, 0.1}

\graphicspath{
{figures/} %
}

\usepackage[nameinlink,capitalise]{cleveref}
\crefname{enumi}{}{}  %
\crefrangeformat{figure}{Figures~#3#1#4--#5#2#6}

\AtBeginDocument{%
  \providecommand\BibTeX{{%
    \normalfont B\kern-0.5em{\scshape i\kern-0.25em b}\kern-0.8em\TeX}}}

\copyrightyear{2022}
\acmYear{2022}
\setcopyright{acmcopyright}\acmConference[UIST '22]{The 35th Annual ACM Symposium on User Interface Software and Technology}{October 29-November 2, 2022}{Bend, OR, USA}
\acmBooktitle{The 35th Annual ACM Symposium on User Interface Software and Technology (UIST '22), October 29-November 2, 2022, Bend, OR, USA}
\acmPrice{15.00}
\acmDOI{10.1145/3526113.3545652}
\acmISBN{978-1-4503-9320-1/22/10}

\begin{document}

\title[Fibercuit]{Fibercuit: Prototyping High-Resolution Flexible and Kirigami Circuits with a Fiber Laser Engraver}

\author{Zeyu Yan}
\authornote{Both authors contributed equally to this research.}
\email{zeyuy@umd.edu}
\affiliation{%
  \institution{Department of Computer Science, University of Maryland, College Park}
  \streetaddress{8125 Paint Branch Dr.}
  \city{College Park}
  \state{MD}
  \country{USA}
  \postcode{20740}
}

\author{Anup Sathya}
\authornotemark[1]
\email{anupsat@umd.edu}
\affiliation{%
  \institution{Department of Computer Science, University of Maryland, College Park}
  \streetaddress{8125 Paint Branch Dr.}
  \city{College Park}
  \state{MD}
  \country{USA}
  \postcode{20740}
}
\author{Sahra Yusuf}
\email{fyusuf4@gmu.edu}
\affiliation{%
  \institution{George Mason University}
  \city{Fairfax}
  \state{VA}
  \country{USA}
}

\author{Jyh-Ming Lien}
\email{jmlien@gmu.edu}
\affiliation{%
  \institution{George Mason University}
  \city{Fairfax}
  \state{VA}
  \country{USA}
}

\author{Huaishu Peng}
\email{huaishu@umd.edu}
\affiliation{%
  \institution{Department of Computer Science, University of Maryland, College Park}
  \streetaddress{8125 Paint Branch Dr.}
  \city{College Park}
  \state{MD}
  \country{USA}
  \postcode{20740}
}

\renewcommand{\shortauthors}{Yan and Sathya, et al.}

\begin{abstract}

Prototyping compact devices with unique form factors often requires the PCB manufacturing process to be outsourced, which can be expensive and time-consuming.
In this paper, we present Fibercuit, a set of rapid prototyping techniques to fabricate high-resolution, flexible circuits on-demand using a fiber laser engraver. 
We showcase techniques that can laser cut copper-based composites to form fine-pitch conductive traces, laser fold copper substrates that can form kirigami structures, and laser solder surface-mount electrical components using off-the-shelf soldering pastes. 
Combined with our software pipeline, an end user can design and fabricate flexible circuits which are dual-layer and three-dimensional, thereby exhibiting a wide range of form factors. 
We demonstrate Fibercuit by showcasing a set of examples, including a custom dice, flex cables, custom end-stop switches, electromagnetic coils, LED earrings and a circuit in the form of kirigami crane.
\end{abstract}



\begin{teaserfigure}
  \includegraphics[width=\textwidth]{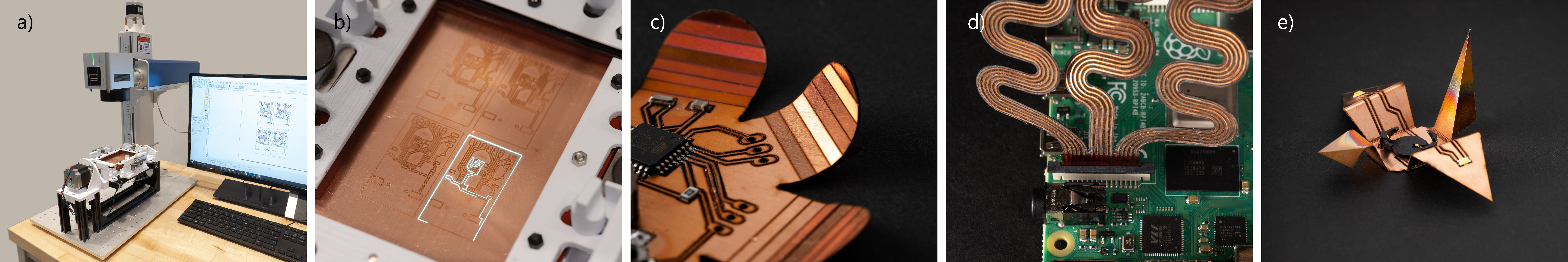}
  \caption{Fibercuit overview. a) Our fiber laser engraver setup with a custom rotary table. b) Cutting the conductive layer on our custom composite using the fiber laser. c) A kirigami flower circuit that can function as an Arduino. d) A custom stretchable flex cable. e) A kirigami crane with a battery holder and 2 LEDs.}
  \label{fig:teaser}
\end{teaserfigure}

\maketitle

\section{Introduction}

In the ongoing effort to weave computing into our daily lives~\cite{weiser1991computer}, designing smart devices has been a major research focus in HCI. As these devices aim to discreetly recede into the environment or onto the user's body, they usually possess small and unconventional form factors. Building and prototyping such devices requires multiple rounds of iterative design to successfully integrate the function into the form. While the consumerization of 3D printers and laser cutters has shortened the iteration cycle, this advancement has been largely withheld from the PCB design and manufacturing process. Due to their flexibility and convenience, breadboards are still the most popular method to quickly prototype circuits. Apart from requiring bulky breakout boards for all surface-mounted components (SMD), integrating breadboards into compact designs where the form factor ideally blends into the background --- like IoT and wearables --- remains a challenge. In these instances, researchers and makers rely on outsourcing PCB fabrication to a small-batch manufacturer. As this outsourcing is not just limited to the final product but also for intermediate iterations, along with the economic burden, this introduces weeks to months of delay in the prototyping process.

Thus far, chemical etching~\cite{wikihow_2021_create} and CNC milling ~\cite{LPKF,bantam_tools} have been two of the most commonly used circuit prototyping methods by makers and researchers. 
In chemical etching, copper is selectively dissolved on a masked substrate to create conductive traces. This method can be used to create circuits that are either rigid or flexible. Unfortunately, the chemical etching process can easily lead to over or under etching of the traces. The resolution, and therefore the minimum width of the traces and pads, is often fairly large~\cite{unimorph}, limiting the capability to prototype small circuits or those which require fine-pitch SMD components. 
Handling corrosive and toxic chemicals in a maker environment also presents a few safety challenges. 
Desktop CNC milling machines have been used to manufacture rigid PCBs on solid substrates. Although reliable and safe, this method cannot always support unconventional and compact form factors which require a flexible PCB. 
Circuits have also been directly integrated into paper, thermoplastic, and laser-cut artifacts using conductive ink~\cite{siegel2010foldable, kawahara2013instant, nisser2021laserfactory,MorphingCircuit}, filament~\cite{ThermoformedCircuitBoards, gong2021metasense} or copper wires~\cite{peng20163delectromagnetic}. Although these methods afford unconventional shapes, like chemical etching, the electrical characteristics are constrained by the methods and the materials. For example, soldering small, fine-pitched SMD components onto a paper-based circuit or a 3D printed object is extremely challenging, if not impossible.

This paper presents Fibercuit, a suite of techniques to fabricate high-resolution flexible circuits suitable for rapid prototyping. 
Our techniques are centered around a desktop fiber laser engraver.
These machines are similar to $CO_2$ laser cutters which are commonly seen in maker spaces or lab settings, but are usually used to engrave metallic materials, \textit{i.e.}, engraving initials on the back of a mobile phone or a tablet.
In this work, we re-purpose the fiber laser engraver to fabricate custom flexible circuits. 
We showcase the laser cutting workflow to fabricate standard flexible circuits, as well as a suite of additional techniques using the fiber laser engraver to solder SMD components and fabricate circuits with kirigami shapes.
These techniques are unique to fiber laser machines and, when combined, allow us to fabricate circuits with greater varieties in size, stiffness, and malleability; The copper traces made with Fibercuit are also high-resolution (as low as 8 mils), which, to the best of our knowledge, was typically achieved only in professional workflows in the past.

We report the configuration of the machine, the process of preparing custom circuit composites, and the techniques we have developed to enable the fabrication of flexible circuit boards.
We also introduce a software interface that allows an end user to design custom circuits on kirigami shapes by simulating and visualizing the folding process. The software also generates the vector files required by the fiber laser engraver to fabricate the kirigami circuit.
We conclude with several fabricated examples that highlight the benefits of our method, along with a discussion of the limitations of our approach and possible future improvements.

\section{Related Works}
Our contribution builds upon prior research focused on fabricating circuits using off-the-shelf machines, fabricating circuits using a laser beam, and fabricating circuits beyond conventional 2D form factors.

\subsection{Fabricating Circuits Using Off-the-Shelf Machines}

As discussed in the introduction, several approaches have been adopted to fabricate circuits on demand. For example, CNC machines such as LPKF ProMat~\cite{LPKF} and Bantam PCB milling tools ~\cite{bantam_tools} have been used in maker spaces and lab settings as they offer reliable creation of conductive traces. However, because these machines use a high-speed spindle to selectively remove copper from traditional FR-4 blank boards, the substrate must be rigid. Flexible circuits that are commonly used in small devices or wearables cannot be fabricated with most CNC machines.

To make flexible circuits in-house, researchers have proposed different fabrication techniques utilizing off-the-shelf machines or custom circuit substrates. 
For example, chemical etching~\cite{wikihow_2021_create} is a widely adopted method in the DIY community~\cite{etching_flex_circuits} as well as in HCI research (\textit{e.g.},~\cite{unimorph}) as the procedure does not require any highly specialized equipment. 
However, the chemical reaction cannot guarantee high-quality conductive traces as over-etching and under-etching are very common. 
Handling corrosive chemicals also requires additional safety measures. 
Researchers have also explored the fabrication of flexible circuits using low-cost plotting tools. For example, vinyl cutters have been used to cut conformal conductive traces out of adhesive foils~\cite{midas} and gold leaf~\cite{kao2016duoskin}. 
A series of works ~\cite{printscreen, gong2014printsense, kawahara2013instant, groeger2018objectskin, perumal2015printem, cheng2020silver, olberding2015foldio} present techniques that utilize commodity inkjet printers to fabricate circuits on paper or custom film. 
While these approaches can create flexible circuits on thin-film substrates, the conductive traces are either low-resolution (\textit{e.g.} with a width limit of 0.5mm or 20 mils, according to \cite{kawahara2013instant}), or have low conductivity, due to the ink's high volume resistance compared to that of copper (4-\SI{200}{\micro\ohm\per\centi\meter} vs. \SI{1.68}{\micro\ohm\per\centi\meter}). Substrate materials such as paper are also fragile, requiring extra care during prototyping. 

Similar to the aforementioned approaches, Fibercuit is based on using off-the-shelf machinery with a custom substrate. However, since our method removes copper material with a high-energy laser beam that has no direct contact with the substrate, Fibercuit can produce high-resolution conductive traces suitable for prototyping small and flexible circuits. Aside from this, our custom substrate is based on Kapton tape, providing higher robustness when compared to other flexible substrates such as paper.

\subsection{Fabricating Circuits using a Laser Beam}

Due to the democratization of $CO_2$ gas laser cutters, they are now commonplace in research labs and maker spaces. This has inspired HCI researchers to explore methods to fabricate circuits with lasers. Since $CO_2$ lasers are primarily used to cut organic materials (\textit{e.g.} acrylic, rubber, and wood) and not metals, they are unable to cut conductive traces directly. 
Instead, researchers have employed laser cutters to create stencils, masks, or molding templates into which conductive material is added. 
For example, Silicone Devices~\cite{nagels2018silicone} presents a DIY workflow to fabricate multi-layered soft circuits.
The $CO_2$ laser precisely cuts layered vinyl masks into which liquid metal can be injected.
Similarly, iSkin~\cite{weigel2015iskin} laser-patterns silicone elastomers such as PDMS and cPDMS to create conductive lines and electrodes, onto which additional silicone is added to seal the circuit traces.

Recently, researchers have also explored the possibility of directly generating conductive traces using a $CO_2$ laser cutter.
LaserFactory~\cite{nisser2021laserfactory} demonstrates a modified end effector that can deposit and solidify conductive silver ink using laser-generated heat.
CircWood~\cite{wooden_circuit} presents a novel approach to generate electrical traces by partially carbonizing the surface of the wood using a laser cutter. LASEC~\cite{lasec} uses a laser cutter to ablate conductive materials such as ITO, Ag paint and PEDOT:PSS to form stretchable conductive traces. 
While these techniques open up opportunities to enable the rapid prototyping of circuits with common laser machinery, the circuit size and the trace width are generally incomparable with flexible printed circuits from the industry, and thus, circuits fabricated using the aforementioned methods are usually sizeable.

Besides commodity $CO_2$ laser machines, high-end laser machines have been used to prototype flexible circuits with small form factors. For example, UV laser machines that specialize in circuit prototyping have been used to manufacture flexible wearable electronic devices (\cite{material_softElectronics, electroDermis}). Advanced laser machines that include both $CO_2$ and fiber laser heads (\textit{e.g.} ~\cite{fusion_pro}) have been used to fabricate flexible circuits ~\cite{multi_layer_flex_pcb}. 

Fibercuit uses a commodity fiber laser engraver machine that contains a single fiber laser source. We showcase how Fibercuit can cut substrates which contain both a copper layer as well as a polyamide Kapton layer with fine details. Since a fiber laser engraver is affordable (as of 2022, a 50W desktop fiber laser engraver~\cite{vmadecnc} is around 5K USD, which is within the same price range as a hobby $CO_2$ laser such as ~\cite{glowforge}), our approach has the potential to democratize ways of rapid prototyping high-quality flexible circuits in lab and maker settings.

\subsection{Fabricating Circuits beyond 2D}

In order to add sensing and display capabilities to daily objects or environments, HCI researchers have proposed different ways to blend circuits into ambient space through hydrography~\cite{groeger2018objectskin}, inkjet printing~\cite{printscreen, kawahara2013instant}, spray painting~\cite{wessely2020sprayable, zhang2017electrick}, and custom plotting devices~\cite{pourjafarian2022print, cheng2021duco}. 
These approaches allow post-hoc application of in-situ circuitry on existing objects or surfaces.

Researchers have also looked into prototyping circuits in 3D form factors, commonly through 3D printing. For example, custom sensors can be printed either directly using conductive filament~\cite{capricate, oh_snap, gong2021metasense} or by post-treating the materials using laser fusing~\cite{fiberwire} or electroplating ~\cite{electroplating,ThermoformedCircuitBoards}. 
Alongside multi-material printing, conformal circuits or breadboard structures can be integrated into 3D printed objects with conductive ink injection~\cite{curveBoards, seriesoftubes} or thermal mounted copper foils~\cite{surfcuit}. 

Compared to 3D printing the entire hull of the object, quasi-kirigami methods like cutting, folding and bending can create 2.5D circuits with flat materials.
As examples, Lo et.al~\cite{shrinkycircuits} and Wang et.al\cite{MorphingCircuit} manage to create 2.5D circuits from pre-processed 2D materials. ~\cite{midas, olberding2015foldio, siegel2010foldable, microcontrollersasmaterial} combine printed circuits and the flexibility of paper to customize paper crafts with electronics.

While Fibercuit is primarily a rapid prototyping method for small-scale flexible circuits, we also present additional techniques to fold copper substrates to form circuits beyond 2D. Unlike previous work that uses $CO_2$ laser cutters to bend acrylic sheets based on gravity~\cite{laserorigami, laser_kirigami_vision, laser_freely_moving,bending_magnetic}, a fiber laser can control the bending angle of a metallic sheet computationally ~\cite{laser_freely_moving, laser_forming_origami, hao2021metal}. We showcase how the bending of metal sheets can be combined with circuit prototyping to create 3D artifacts with integrated mechanical and electrical features. %

\section{Background on Fiber Laser}\label{background}

$CO_2$ lasers are now commonplace in a lot of maker environments, but the commercial expansion and adoption of fiber lasers are relatively recent and limited. 
For a user, the fundamental difference between a $CO_2$ laser and a fiber laser is the type of materials that can be processed by each type of laser. 
$CO_2$ lasers (\SI{10600}{\nano\meter} range) are able to cut organic materials like textiles, wood and cardboard. 
For inorganic materials like metal, a high-power fiber laser (\SI{1060}{\nano\meter} range) is appropriate. 
In addition to the wavelength, fiber lasers innately produce a higher quality (better focused) beam which not only increases the efficiency of the laser but also enables a higher resolution while cutting and rastering. 
Fiber lasers also use a 2-axis galvanometer laser scanner as opposed to the slower 2-axis gantry system commonly seen in $CO_2$ lasers. This results in a smaller working area, but much faster cutting and rastering speeds.

While the most popular consumer demand for fiber lasers revolve around engraving metallic devices to customize them, we see an opportunity to re-purpose a fiber laser engraver for flexible circuit prototyping due to its high-efficiency, high-quality cutting results and high speed. Our work and experiments are based on an off-the-shelf 50W desktop fiber laser engraver ~\cite{vmadecnc} as is shown in Figure ~\ref{fig:teaser}a.

\section{Fibercuit}

\begin{figure}[htb!]
\centering
  \includegraphics[width=\columnwidth]{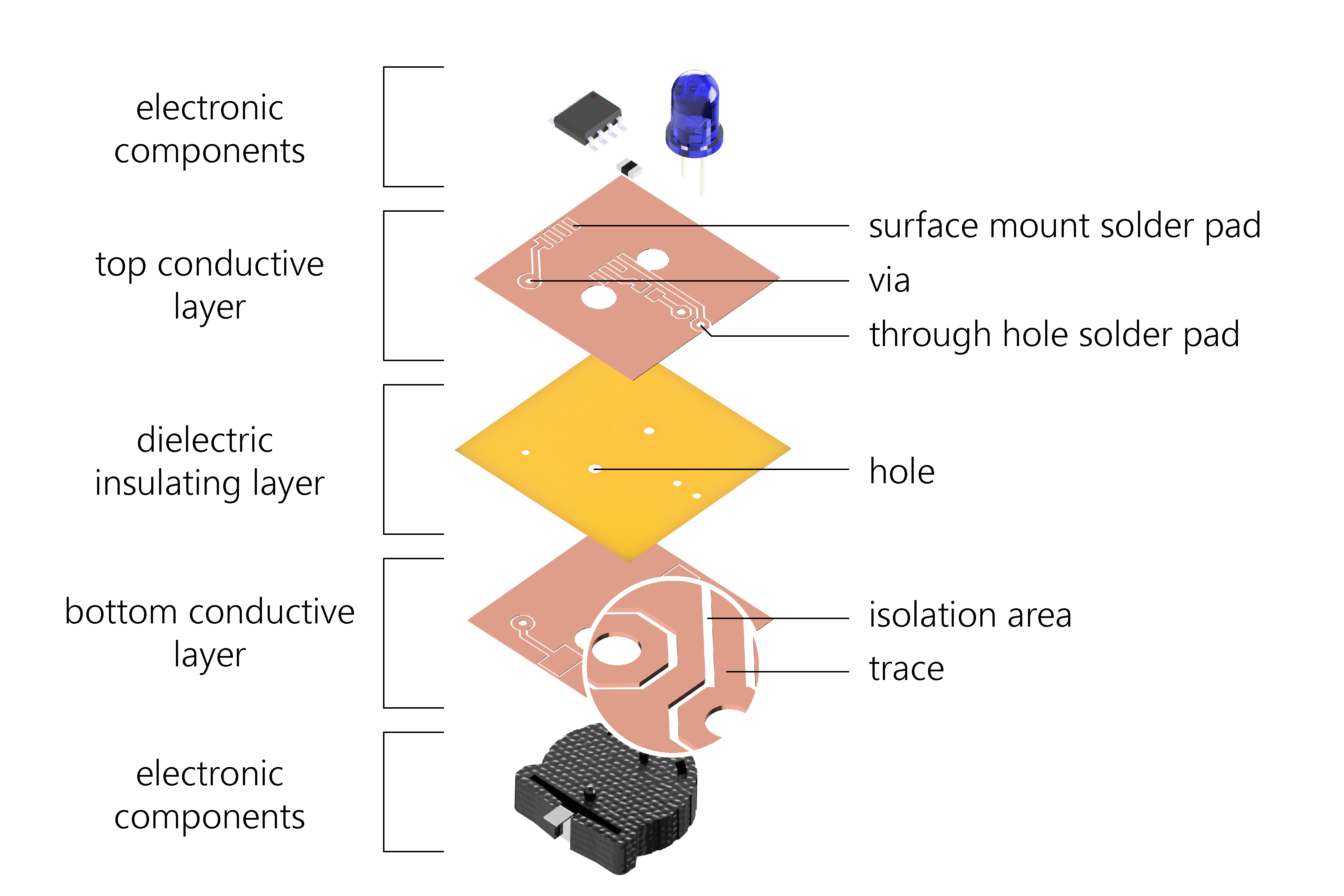}
  \caption{An illustration explaining the structure of a circuit.}~\label{fig:PCBStructure}
\end{figure}

In essence, a basic dual-layer flexible circuit is a sandwich structure with a conductive layer on either side and a dielectric insulating layer in between. The top and bottom layers contain conductive traces, pads, and pours. The dielectric layer hosts the conductive elements and electrically isolates the two layers (Figure ~\ref{fig:PCBStructure}).
In addition to the sandwich structure, a circuit board must perform three functions. \textit{One}, enable the affixing of electronic components at designated spots  --- through-hole or surface mount. \textit{Two}, provide reliable electrical contact between the terminals of the electronic components --- traces. \textit{Three}, allow interconnections between conductive layers --- vias. 

As a dual-layer circuit is an extension of a single-layer circuit---with an additional conductive layer on the opposite side and vias to interconnect between layers, we first explain our basic techniques with a single-layer circuit, then introduce the techniques required to manufacture a dual-layer circuit. We first introduce a custom substrate containing the conductive and dielectric layers that can be cut by the fiber laser. We then introduce techniques to create traces and the isolation between them to form functional circuits. We explain how vias are created to enable dual-layer circuits and finally detail the post-processing methods to add robustness. All laser engraving parameters for these techniques are summarized in Table~\ref{tab:paras}, towards the end of the paper. 

\subsection{Preparing Circuit Substrates} \label{substrate}

Our circuit substrate is a structure containing copper sheets and single or double-sided polyamide Kapton tape. The copper sheet acts as the conductive layer; the Kapton tape acts as the dielectric insulating layer and also structurally holds the traces, pads, and pours on the conductive layer.

\subsubsection{Conductive Layer}
We experimented with copper sheets of different thicknesses (0.03 mm, 0.05 mm, 0.1 mm, 0.15 mm, and 0.2 mm). A 50W fiber laser machine can successfully cut all copper sheets thoroughly. As our primary goal is to fabricate flexible circuits, we excluded thicker copper from our experiments. Besides, using thicker copper introduces trade-offs which are discussed in Section~\ref{thickness}.

\subsubsection{Dielectric Insulating Layer}

When identifying an insulating material, one primary factor is to ensure that it can be cut by a fiber laser engraver. Holes, drills, and the dimension outline of the board require both the copper and the insulating layers to be cut through. As discussed in Section \ref{background}, the wavelength of the fiber laser is not ideal to cut through organic materials. We therefore tested multiple materials, including Kapton tape, thin acrylic sheet, Delrin, and styrene, to find one that can be reliably cut with a fiber laser source.
Among all the materials, polyamide Kapton was the most suitable. It can be lightly engraved by a fiber laser but not cut through. When bonded to copper sheets, however, the heat generated while cutting the copper transfers to Kapton, allowing it to be cut through. Kapton also possesses a great dielectric strength (7700 V/mil for a 1 mil thick film) and remains stable from -269 to \SI{400}{\celsius}. We experimented with Kapton tapes of 1 mil, 2 mils, and 5 mils thickness which can all be reliably cut with the fiber laser when bonded with a copper sheet.

\begin{figure}[htb!]
\centering
  \includegraphics[width=\columnwidth]{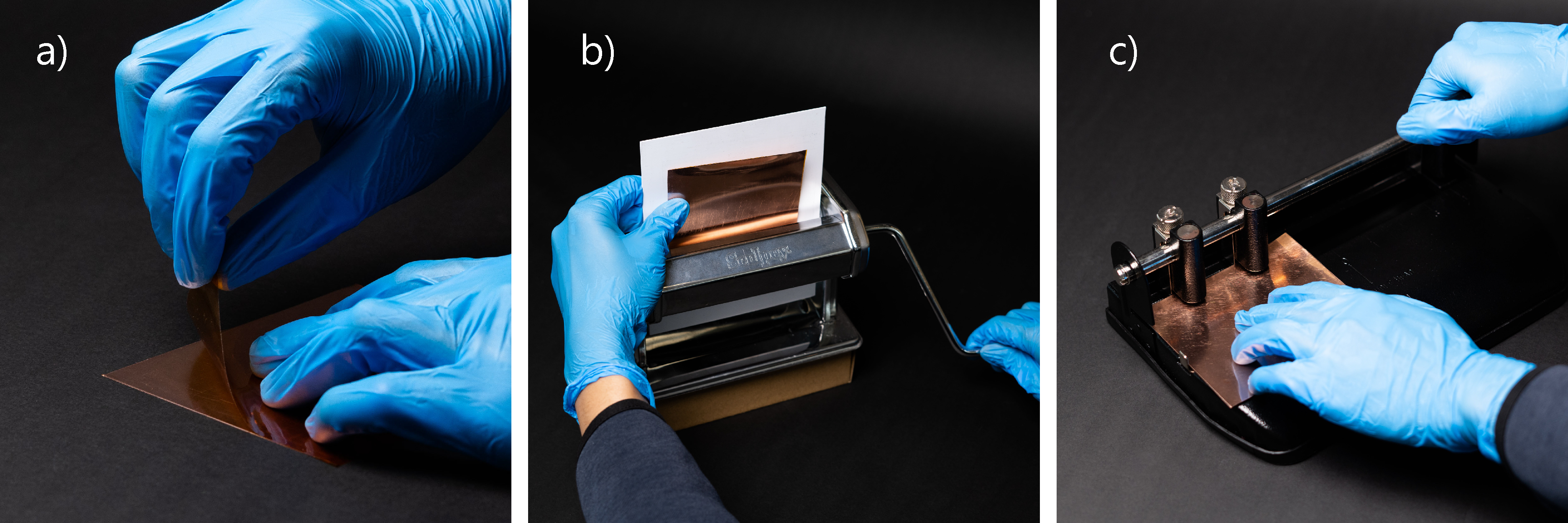}
  \caption{Circuit substrate preparation. a) Applying Kapton tape to the copper sheet. b) Passing the copper-Kapton composite through a roll former. c) Punching alignment holes. }~\label{fig:composite}
\end{figure}

\subsubsection{Fabrication}
As the composite material is directly processed by the laser, it is imperative to ensure a strong bond between the copper and Kapton layers. 
We found that the silicone adhesive on the Kapton tape performs satisfactorily.
It is also important to avoid bubbles between the two layers, as failures in this step will result in an uneven substrate surface that affects the laser cutting process.
The procedure described below highlights the steps required to ensure adequate adhesion.

\textit{Step 1}: The copper sheet needs to be thoroughly cleaned with Isopropyl Alcohol (IPA) to remove surface dirt and grime. 
In case the copper sheet is heavily oxidized, a light sanding with fine-grit sandpaper or an abrasive pad can also be employed. Wearing gloves while handling the copper sheet prevents further oxidization or grime collection. 

\textit{Step 2}: Once the surface of the copper is thoroughly cleaned, fresh Kapton tape is applied (Figure \ref{fig:composite}a). 
The Kapton-taped copper sheet is passed through the slip of a roll former (Figure \ref{fig:composite}b) together with a piece of styrene to evenly press the composite together.

\textit{Step 3}: On occasions where the substrate needs to be intermittently taken on and off the laser bed during the cutting process, we create alignment holes in the substrate using a hole punch tool that matches the dowel pins on our material clamping bed (Figure \ref{fig:composite}c).

Note that \textit{Step 2} is critical in the material preparation process and should not be ignored. It performs three functions. \textit{One,} it allows the silicone adhesive on the surface of the Kapton tape to strongly adhere to the copper. \textit{Two,} it helps remove any air bubbles that arise during the taping process. \textit{Three,} it flattens the copper-Kapton composite, removing any bends or kinks, creating a flat surface for the laser engraver to work effectively. 
The copper-Kapton composite is passed multiple times through the slip roll former (while gradually reducing the gap between the rollers) until the stock is adequately flat, kink-free, and bubble-free.

Also note that there can be slight variations to the composite preparation procedure above. For example, while using Kapton tape that is thinner than 2 mils, ensuring a bubble-free surface is challenging. 
In this case, soapy water is sprayed onto the surface of the copper and the adhesive side of the Kapton tape. 
After applying the Kapton tape to the copper sheet, a scraper is used to squeeze the soapy water out --- thereby pushing the bubbles out. 

While fabricating dual-layer circuits, we follow the same procedure up to this point by replacing the single-sided Kapton with a double-sided Kapton. After one conductive layer is satisfactorily attached to one side of the Kapton, we repeat the process to attach another copper sheet to the other side of the Kapton, forming a sandwich.

\subsection{Fabricating Conductive Traces} \label{cutting}

As shown in Figure \ref{fig:PCBStructure}, a conductive trace is a connection between pads that are bonded to the insulating Kapton layer while being fully isolated from the rest of the conductors. Therefore, creating conductive traces in a circuit board requires the selective removal of copper to prevent an electrical short. It is critical for the fiber laser to penetrate the copper layer while keeping the bottom isolating layer intact to ensure that different traces are not electrically shorted. The copper in the isolating area should also be easily removable without damaging the structure of the circuit.

\subsubsection{Cutting the Outline of the Traces and the Isolating Area.} 
A conductive trace is created using vector cutting, where the beam of the fiber laser irradiates the outline of the surrounding isolating area. In order to gain precise control over the cutting depth to ensure that the laser beam fully penetrates the top copper layer while keeping the bottom layer intact, the vector cutting process is performed with multiple passes at high speed. The cutting parameters vary based on the substrate composition (See Table~\ref{tab:paras} for details). For example, for the sample circuit presented in Figure~\ref{fig:sample}, the conductive traces are cut with the laser at a speed of \SI{500}{\milli\meter\per\second}, a power of 60\%, and a total of 14 passes. Note that excess heat may be generated during the cutting process, creating local warping or affecting the adhesion of the Kapton. We alleviate them by introducing a delay between successive cutting passes, which allows the material to cool down and avoids the adhesive's the carbonization (parameters reported in Table~\ref{tab:paras}). %

After experimenting with multiple trace widths, we found that the minimum trace width that can be cut with $\sim$100\% repeatability is 8 mils, which is currently more or less on par with small-batch manufacturers. We also tested the minimal clearance width that can be generated reliably --- 4 mils. We note that the trace width and clearance can potentially be reduced further. For example, we successfully test cut a circuit with a trace clearance of 2 mils and a trace width of 4 mils. However, while the isolation is still achievable, the main problem with the even smaller clearance is that manually peeling the isolation becomes less practical (See Section~\ref{peeling_area}). Thus, we decided not to explore the limits of the trace width, as we believe that the aforementioned resolution is already suggesting that Fibercuit is suitable for prototyping high-resolution circuits and incorporating small SMD components.  (See Figure~\ref{fig:soldering}e and f for examples).

\subsubsection{Peeling the Isolating Area} \label{peeling_area}
\begin{figure}[htb!]
\centering
  \includegraphics[width=\columnwidth]{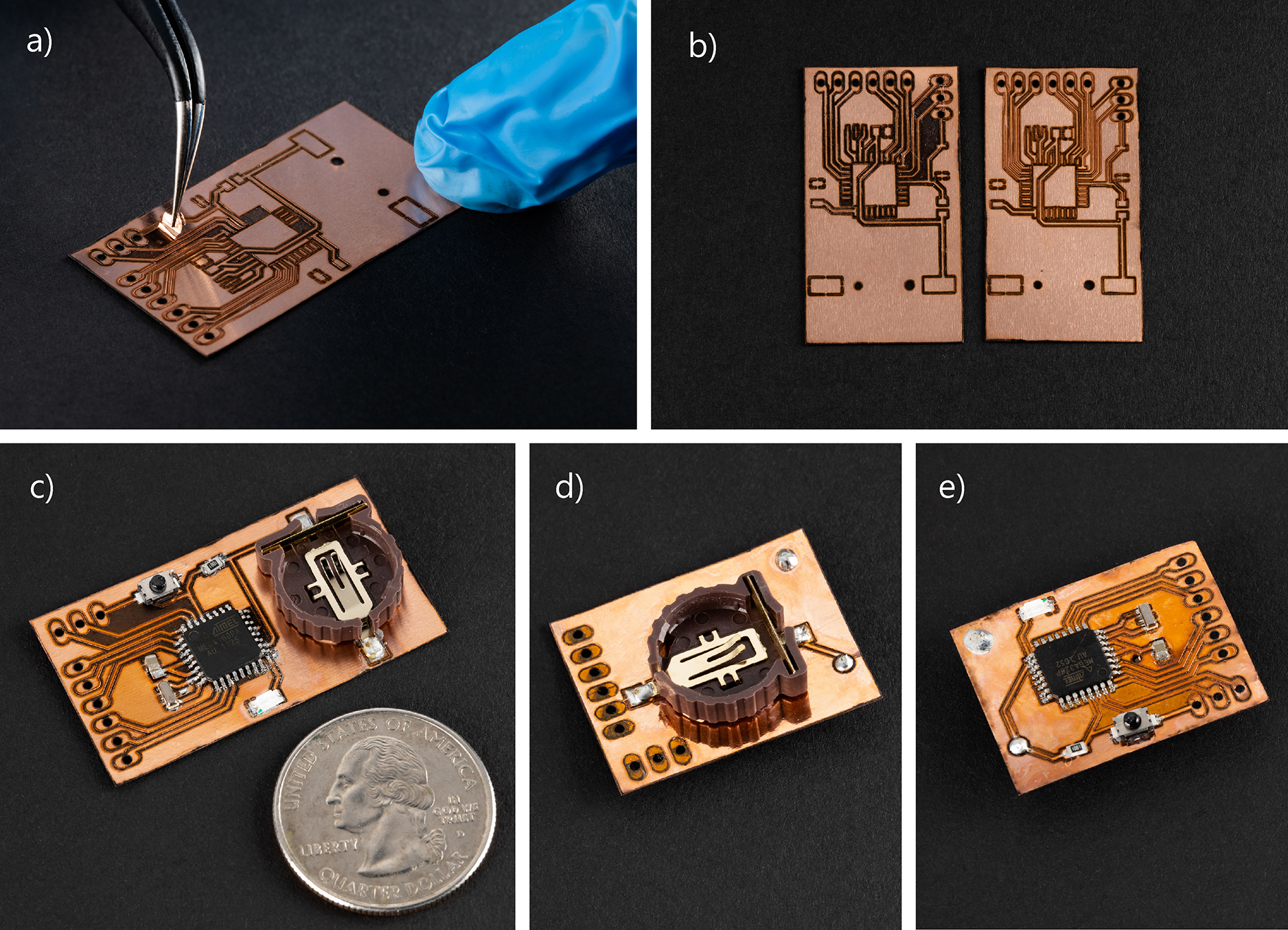}
  \caption{a) Peeling the isolating area using a pair of sharp tipped tweezers. b) Circuit before and after the isolating areas are peeled off. c) Sample circuit in a single-layer format. d) Sample circuit in a dual-layer format (front). e) Sample circuit in a dual-layer format (back).}~\label{fig:sample}
\end{figure}

Once the outline is cut, the isolating area needs to be peeled away from the substrate. This is mostly done manually using a pair of tweezers (Figure \ref{fig:sample}a). To ease the peeling process and to keep the rest of the circuit traces in place, it is helpful to reduce the adhesion between the isolation area and the Kapton underneath. In order to do so, we apply high-speed laser rastering on top of the isolating area. Different from vector cutting, laser rastering scans through the area of interest with the laser beam irradiating an entire area instead of a path. In our case, applying laser rastering to the isolating area with the copper that has to be peeled generates enough heat to slightly carbonate the silicone adhesive of the Kapton tape. It also introduces some internal stress that curls up the edges of the copper. As a result, although a manual process, the isolating area can be easily peeled away from the substrate. Table \ref{tab:paras} reports the rastering parameters that apply to our substrate.

\subsubsection{Cutting Through: Holes, Drills, and Dimension}
Circuit boards usually contain holes and drills to house through-hole components. They require the laser to cut through the conductive layer and the Kapton layer at the same time. The same applies to the board dimension to get the circuit off the working stock. Table~\ref{tab:paras} summarizes the required process parameters. Overall, cutting through needs a much higher energy density at any given time. For example, in the sample circuit shown in Figure~\ref{fig:sample} c, the cutting speed is 5 times slower than that used while cutting the trace outlines. 

\subsubsection{Vias}
As a general design guideline, the via design has to be slightly altered to accommodate our fabrication procedure.
The width of the annular ring is set to 50 mils (\SI{1.27}{\milli\meter}) and the drill size is set to 30 mils (\SI{0.762}{\milli\meter}). Although this is larger than the minimum size of a standard factory manufactured via, the increased drill size allows us to pass a single-core wire through the via. The larger annular ring allows adequate space to solder the inserted wire on both the layers.

\subsection{Circuit Samples} \label{circuitSample}

At this stage, we have the building blocks required to generate both single-layer and dual-layer circuits. 
In Figure~\ref{fig:sample} c, we show a single-layer sample circuit with 7 SMD components, including one 32-pin ATMEGA168 microcontroller (TQFP package) with 0.8 mm pitch, two 0805 SMD resistors and capacitors, one 16MHz crystal oscillator, one 1206 LED, one tactile reset button, and one CR1220 coin cell holder. 
It also has nine through-hole pin-head terminals to provide access to the GPIOs with standard 2.54 mm pitch and 2 mm holes. 
The circuit is laser cut using the copper-Kapton composite with an overall thickness of \SI{0.15}{\milli\meter}. 
The sample circuit board has an overall size of \SI{42}{\milli\meter} by \SI{22.5}{\milli\meter}. 
It can perform standard Arduino Nano functionalities and in a flexible form factor along with an onboard battery. 
The entire fabrication process took approximately 30 minutes, with roughly 5 minutes of laser machining time and 25 minutes of “human time” --- preparing material, peeling the isolation, and soldering the component. As a result, we note that the time taken for a particular circuit has multiple confounding factors, but the “machine time” is often negligible.  %

Figure \ref{fig:sample} d and e demonstrate a dual-layer version of the Arduino Nano circuit. In this version, we move the battery holder to the rear of the circuit. This adds two vias --- one for Vcc and one for ground. The dual-layer circuit board has the same components in a smaller footprint (\SI{30}{\milli\meter} by \SI{22.5}{\milli\meter}).

\subsection{Post Processing} \label{postprocessing}

\begin{figure}[htb!]
\centering
  \includegraphics[width=\columnwidth]{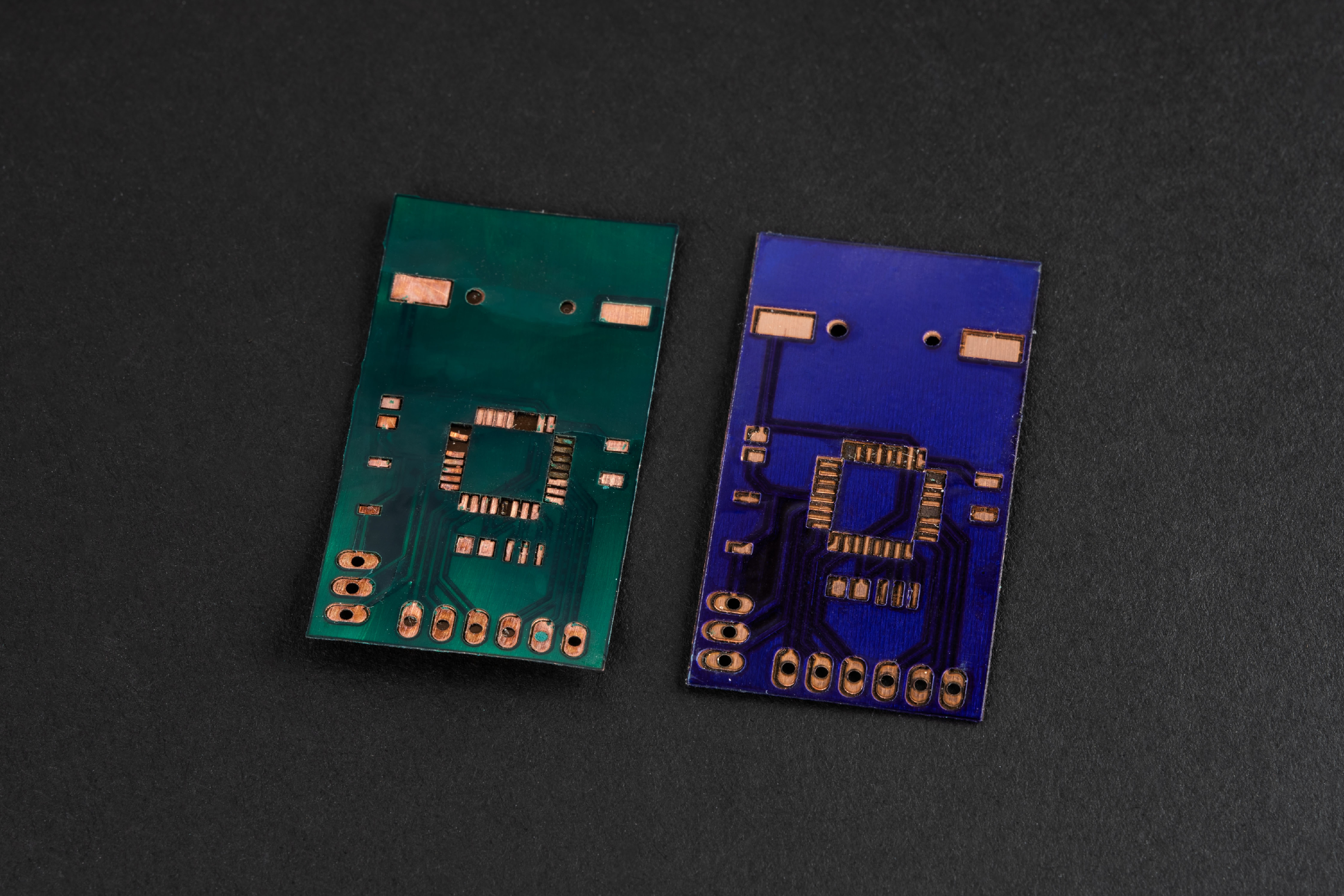}
  \caption{Additional protection can be added to flexible circuits. Left, a UV curable resin is applied to the surface of a single layer circuit. Right, the sample circuit is bonded with a layer of photoresist film.}~\label{fig:mask}
\end{figure}

Although the steps mentioned above allow us to fabricate single- and dual-layer circuits, additional protection can be added to the conductor's surface to prevent oxidization and improve its durability. At this point, a standard UV curable solder mask can be applied to the circuit. Once the solder mask is applied to the entire surface, the circuit can be placed on the laser bed again. The fiber laser can be used in the rastering mode to etch the solder mask away to expose the soldering pads (Figure \ref{fig:mask} left). Alternatively, a layer of pre-cut photoresist film with the solder pads already exposed (Figure \ref{fig:mask} right) can also be applied to the circuit as a protective layer.  

\section{Fibercuit++}
So far, we have introduced the foundations of Fibercuit, a suite of techniques to create conventional single or dual-layer flexible circuits. Beyond that, Fibercuit also affords new electronic prototyping opportunities such as tuning the flexibility of the circuit, using the laser to directly solder the components onto the board, and leveraging copper's malleability to fabricate kirigami circuits that amalgamates electrical and mechanical properties. We now detail these opportunities.

\subsection{Circuits Flexibility} \label{flexibility}

\begin{figure}[htb!]
\centering
  \includegraphics[width=\columnwidth]{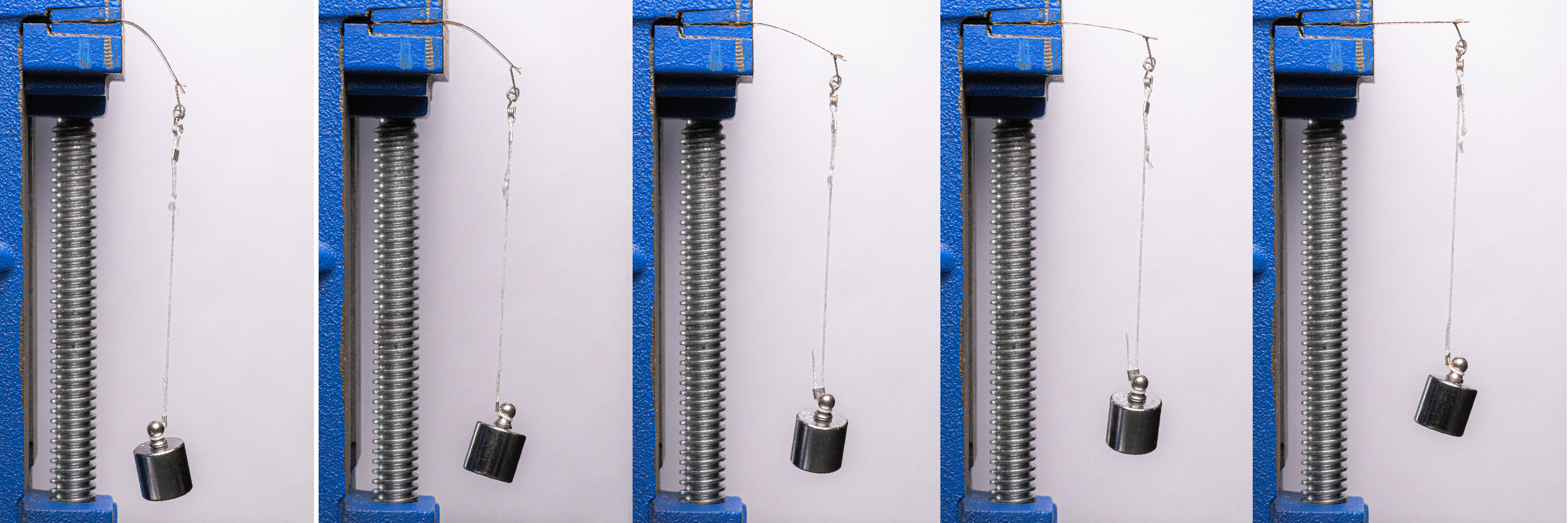}
  \caption{Test samples from left to right: composite with copper layer of 0.03 mm, 0.05 mm, 0.1 mm, 0.15 mm, and 0.2 mm thickness respectively and Kapton layer of 5 mils. Sample size: 20 mm by 10 mm. Weight: 20 g.}~\label{fig:flexibility}
\end{figure}

The copper and the Kapton present their own unique mechanical properties in our dual-material composite. The copper sheet presents plasticity, ductility, and malleability, while the Kapton presents flexibility without much permanent deformation when the applied force is small and spread out. Each material's properties are also heavily dependent on its individual thicknesses. By using and combining a range of thicknesses, we can produce a variety of mechanical properties in the circuit board, ranging from thin and flexible circuits (like a flex-PCB) to thick and rigid circuits. Figure \ref{fig:flexibility} demonstrates the range of rigidity in the composite.

\subsection{Fiber Laser Soldering}

\begin{figure}[htb!]
\centering
  \includegraphics[width=\columnwidth]{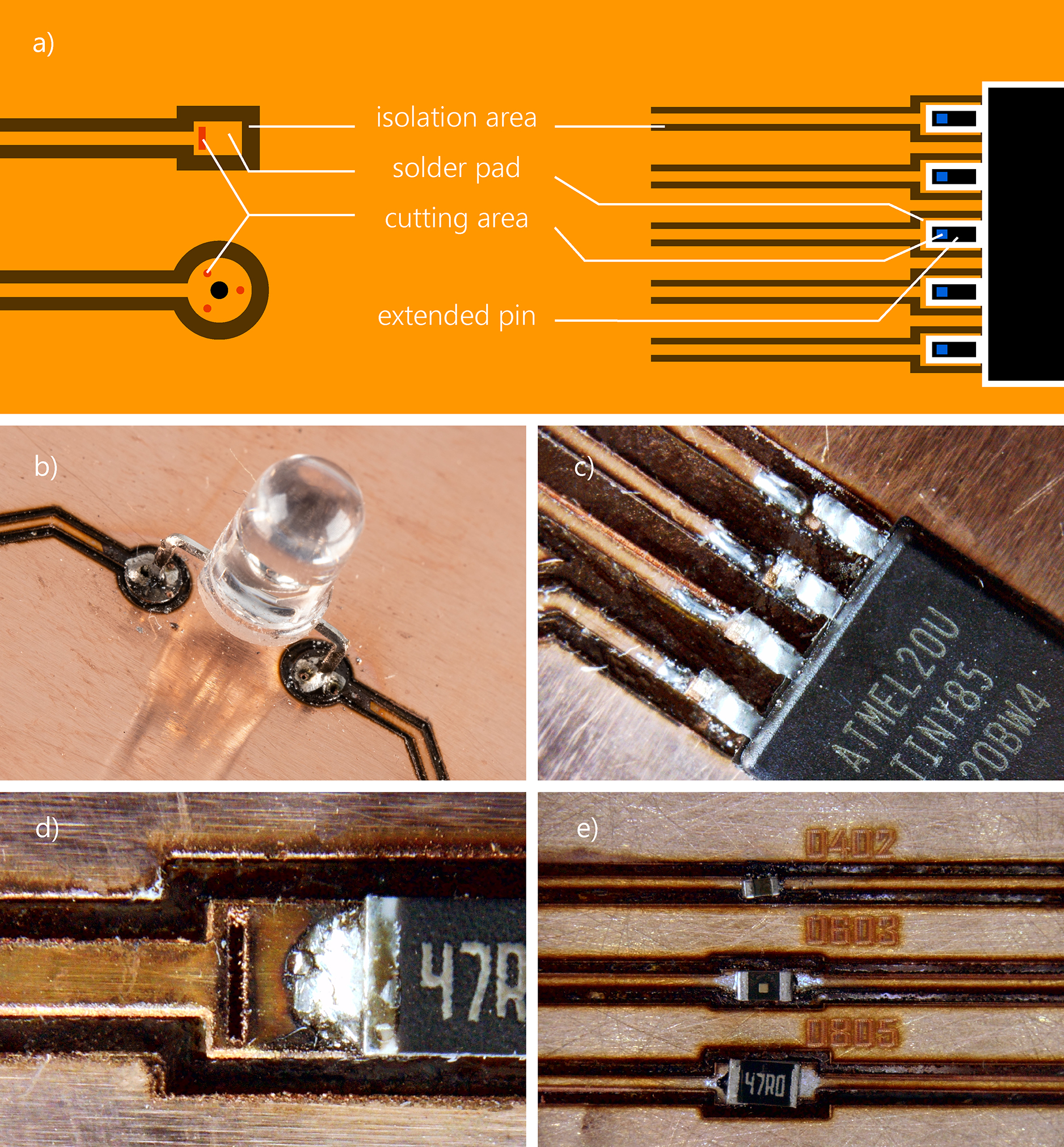}
  \caption{Fiber laser can be used to solder through-hole and SMD components. }~\label{fig:soldering}
\end{figure}

Heat is generated constantly while circuit traces are cut with a fiber laser, which can be potentially utilized to solder electrical components and further mechanize the circuit fabrication process.
Previous work such as ~\cite{nisser2021laserfactory} has shown that by defocusing a $CO_2$ laser, the generated heat can be used to melt and solder conductive inks. Unfortunately, due to copper's high thermal conductivity and the working principle of a fiber laser, high-powered defocused rastering cannot generate enough heat to reflow solder paste. Instead, we propose to cut a small amount of sacrificial material at or around the solder spots to generate the heat required to reflow the solder. 

The general process of fiber laser soldering is to first apply solder paste (\textit{e.g.} \cite{solder_paste}) to the trace terminals, place the components, and then laser cut the surrounding sacrificial area for reflow. While the main soldering principle remains the same, different component packages may require slightly varied approaches.

For through-hole components and vias, there is no need to change the circuit footprints, as they provide existing sacrificial areas (Figure \ref{fig:soldering}a) on the pads,  which can be used to cut a tiny \SI{0.1}{\milli\meter} to \SI{0.3}{\milli\meter} hole (Figure \ref{fig:soldering}c and e). This procedure generates the required heat to reflow the solder and does not affect the electrical or structural integrity of the circuit. 

In other cases, when the footprint does not have a sacrificial area but the component has extended solder pins (like TQFP packages), the laser can directly irradiate the pins (Figure \ref{fig:soldering} b and d). As the laser is calibrated to cut thin copper (\SI{0.03}{\milli\meter}), this generates the required heat to reflow solder but does not damage the extended pins.  

Packages like tubular passive components (capacitors, resistors) do not provide any additional space on the footprint onto which a small hole can be cut. In this case, the footprint of the component is modified during the circuit design phase to enable laser soldering. For example, we slightly extended the pads on a 0805 SMD capacitor (Figure \ref{fig:soldering} e) to create a space where a \SI{0.3}{\milli\meter} hole can be cut to generate the heat required to reflow the solder. Just like the other cases, the Kapton under the 0.3 mm hole remains unscathed.

\subsection{Laser Forming Circuits} \label{laserforming}

By employing the temperature gradient mechanism (TGM) \cite{tgm}, the copper in our composite can be laser formed, \textit{i.e.}, bending upwards and away from the original plane (Figure \ref{fig:bending}a). This can provide additional layers of mechanical freedom, functionality, and interactivity.

\subsubsection{Bending angle vs. pass number}
Essentially, using multiple passes along the same irradiation path, the laser beam induces a steep temperature gradient along the irradiation path across the thickness of the copper sheet.
By performing multiple passes, this process can be used to bend the copper sheet along the path computationally, with the bending angle being estimated directly from the number of cutting passes. 

\begin{figure}[h!]
\centering
  \includegraphics[width=\columnwidth]{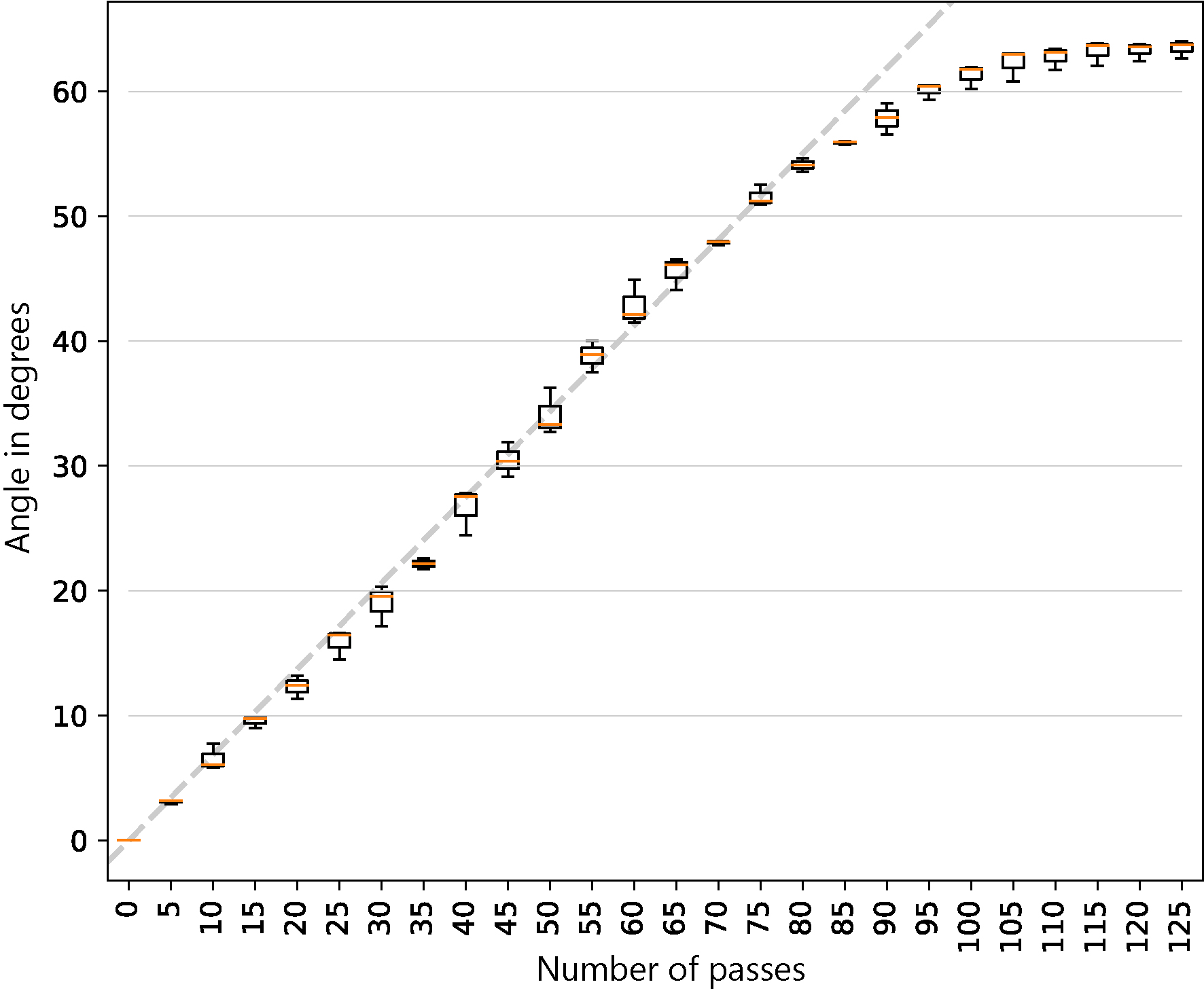}
  \caption{Bending angle vs. engraving pass number.}~\label{fig:curve}
\end{figure}

As can be seen in Figure~\ref{fig:curve}, the bending angle monotonically increases with the number of passes applied along the cutting path, but with a decrease in the bend angle per pass over time. We acquired the number of cutting passes vs. the bending angle relations with the following experiment: we cut samples with the cutting passes increased from 5 to 125 with 5-pass increments and measured the corresponding bending angles. We ran three trials to get the passes vs. angle relations. The Average S.D. across all angles is 0.967 degree. The process parameters are reported in Table~\ref{tab:paras}. 

\subsubsection{Achieve higher bending angle}
\begin{figure}[htb!]
\centering
  \includegraphics[width=\columnwidth]{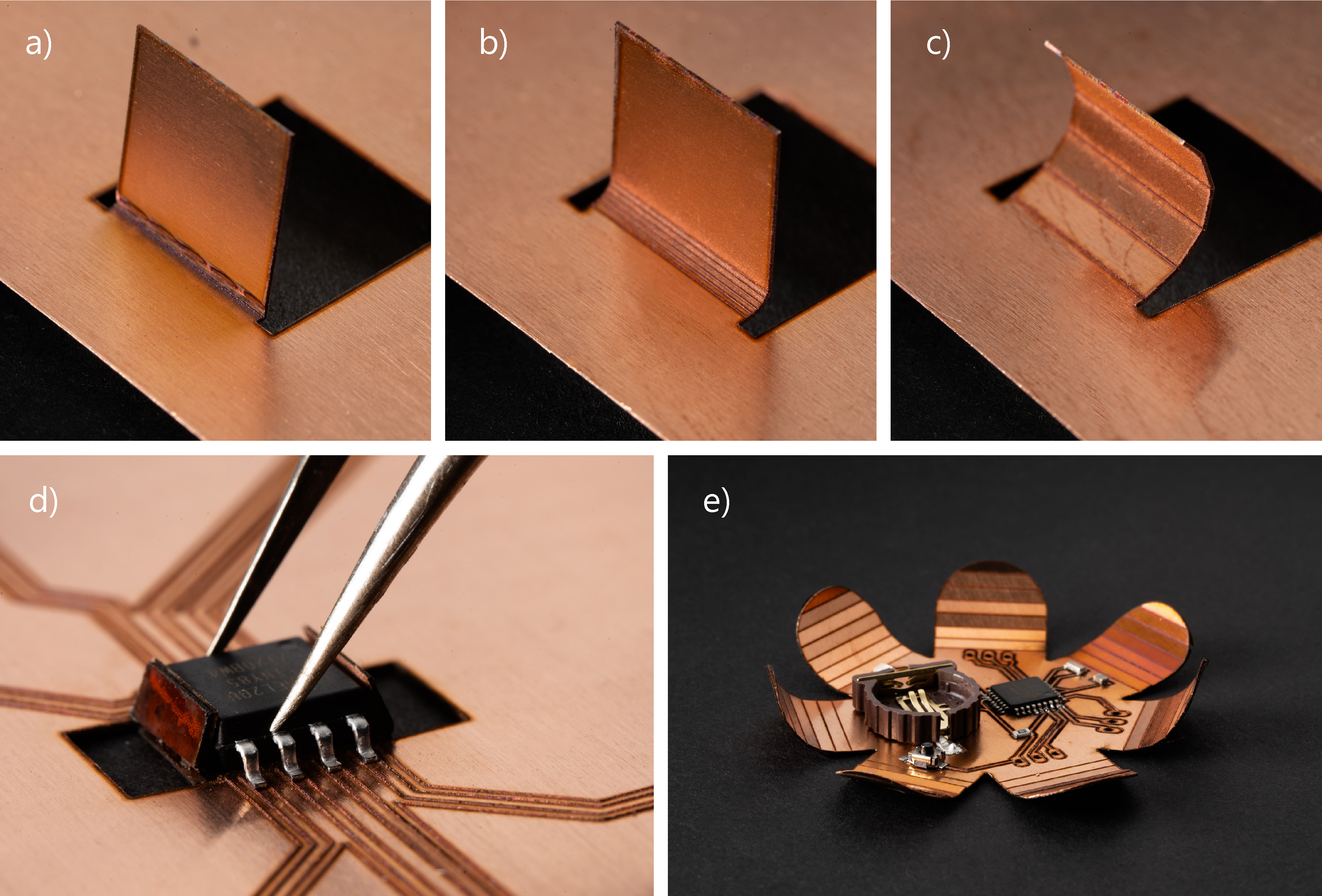}
  \caption{a) to c) Bending copper-Kapton composites to \ang{90} using different approaches. d) A laser formed placement guide for an ATTiny85. e) Arduino Nano circuit on a kirigami flower shape. }~\label{fig:bending}
\end{figure}

As is shown in Figure~\ref{fig:curve}, the relationship between bend angle and number of passes tapers off under high pass numbers. 
To achieve higher bending angle, we have experimented with two empirical solutions: 1) the power of the laser can be mildly increased when the bend angle starts to level out, and 2) create multiple parallel irradiation paths to create multiple bends close to each other. Usually, the bend angle starts to level out at around 40-90 passes (depending on the power percentage setting) when the power is kept constant. 
When it starts to level out, slightly increasing the power by about 3\% will maintain the bend angle per pass. 
Using this constant increase, starting from about 42\% power and going all the way up to 80-85\%, we can increase the ceiling of the bending angle from $\sim$\ang{60} degree to $\sim$\ang{90} (Figure \ref{fig:bending} a) for a given set of process parameters. 
The second approach involves creating multiple closely parallel irradiation paths. 
In this case, the $\sim$\ang{90} bend angle can be achieved while maintaining constant power (Figure \ref{fig:bending} b). 

Sometimes, the Kapton layer interferes with the laser forming process. 
In order to reduce this interference, we flipped the circuit on the bed and selectively etch the Kapton away in rastering mode. 
This removes a small patch of Kapton just under the area that is to be laser formed, therefore reducing the interference of the Kapton layer during laser forming.

\subsubsection{kirigami circuits with laser forming}
This ability to laser form the material allows us to explore unconventional shapes in circuits manufactured using our method. For example, our sample circuit can move away from a traditional rectangular flat shape into a kirigami flower-like shape with the petals bent by the laser (Figure \ref{fig:bending}e). Apart from visual exploration, this technique also allows us to incorporate additional functionality. For example, as shown in Figure \ref{fig:bending}d, the copper can be bent to form a placement guide for SMD components on the circuit board. This aids the user as well as keeps the component in place during the reflow process if a hot air gun is being used. This ability can also afford the fabrication of custom sensors (Section \ref{endstop}) and kirigami objects (Section \ref{crane}). The laser forming not only applies to the circuit's structure but also applies to the traces. This allows us to place components on different planes in a kirigami circuit (See examples in Section \ref{dice} \& \ref{crane}).

\subsection{Order of Operations Involving Laser Soldering and Forming}

Based on the features being incorporated into the circuit board, the order of operations varies. 
Intrinsically, the composite board stock is generated at the very beginning, followed by cutting the pads and conductors, isolating them and then finally soldering the components. 
In cases where laser forming is being incorporated, the order of the forming step is design-dependent. 
In some cases, it's better to solder the components before attempting laser forming as the soldering difficulty increases when the form factor is unconventional.
In other cases the soldering procedure might not be impeded by the change in the form factor; the bending can occur before.
To aid the decision on the operation angle, we have developed a software interface with a 3D view to simulate the circuit before and after laser forming. This is further described in Section \ref{ui}.

\section{Examples}
Fibercuit can be used to fabricate high-resolution, flexible, and kirigami circuits. In this section, we highlight a few archetypes to demonstrate the capabilities and opportunities that Fibercuit enables. 

\subsection{Custom Dice} \label{dice}

\begin{figure}[h]
\centering 
  \includegraphics[width=\columnwidth]{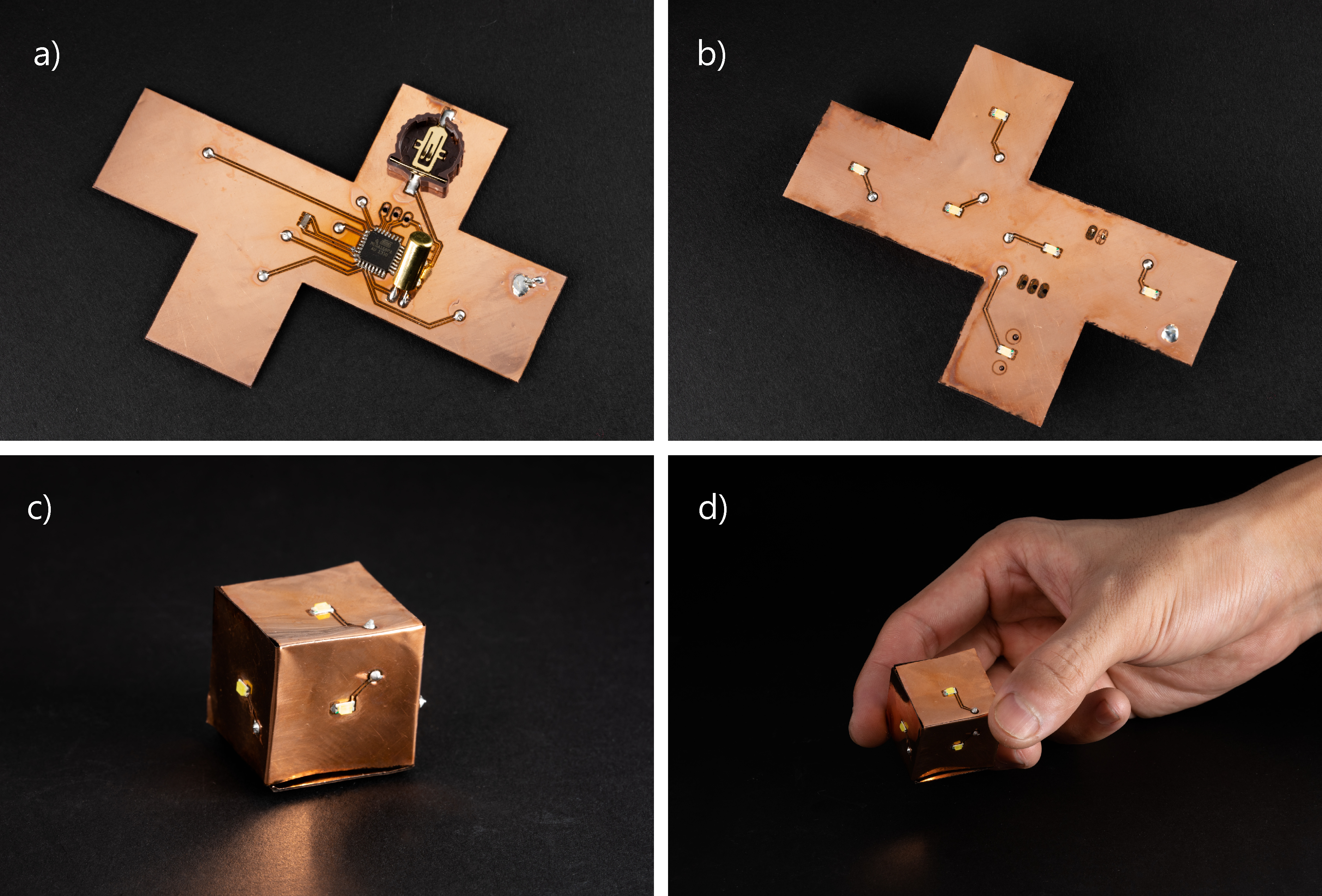}
\caption{A custom dice fabricated using Fibercuit. a) Interior face of the circuit board after the fabrication process. b) Exterior face. c) \& d) Folding the dice and using the circuit as the structure.}
\label{fig:dice}
\end{figure}

As explained in Section \ref{flexibility}, having the ability to play with the malleability of the circuit allows us to prototype unique examples to leverage this property. In this example, we designed a custom dice with LEDs on all 6 sides that are randomly illuminated when the user shakes the dice (Figure \ref{fig:dice}d). The dice contains an ATMega168 (and its operating circuit), a tilt ball switch, a battery holder, and 6 1206 LEDs (Figure \ref{fig:dice}a and b). In a dual-layer design, the LEDs are placed on the exterior face of the dice and the other components are on the interior face. By leveraging the malleability of the circuit, the structural component of our object comes from the circuit itself. 

\subsection{Flex Cable}
\begin{figure}[h]
\centering 
  \includegraphics[width=\columnwidth]{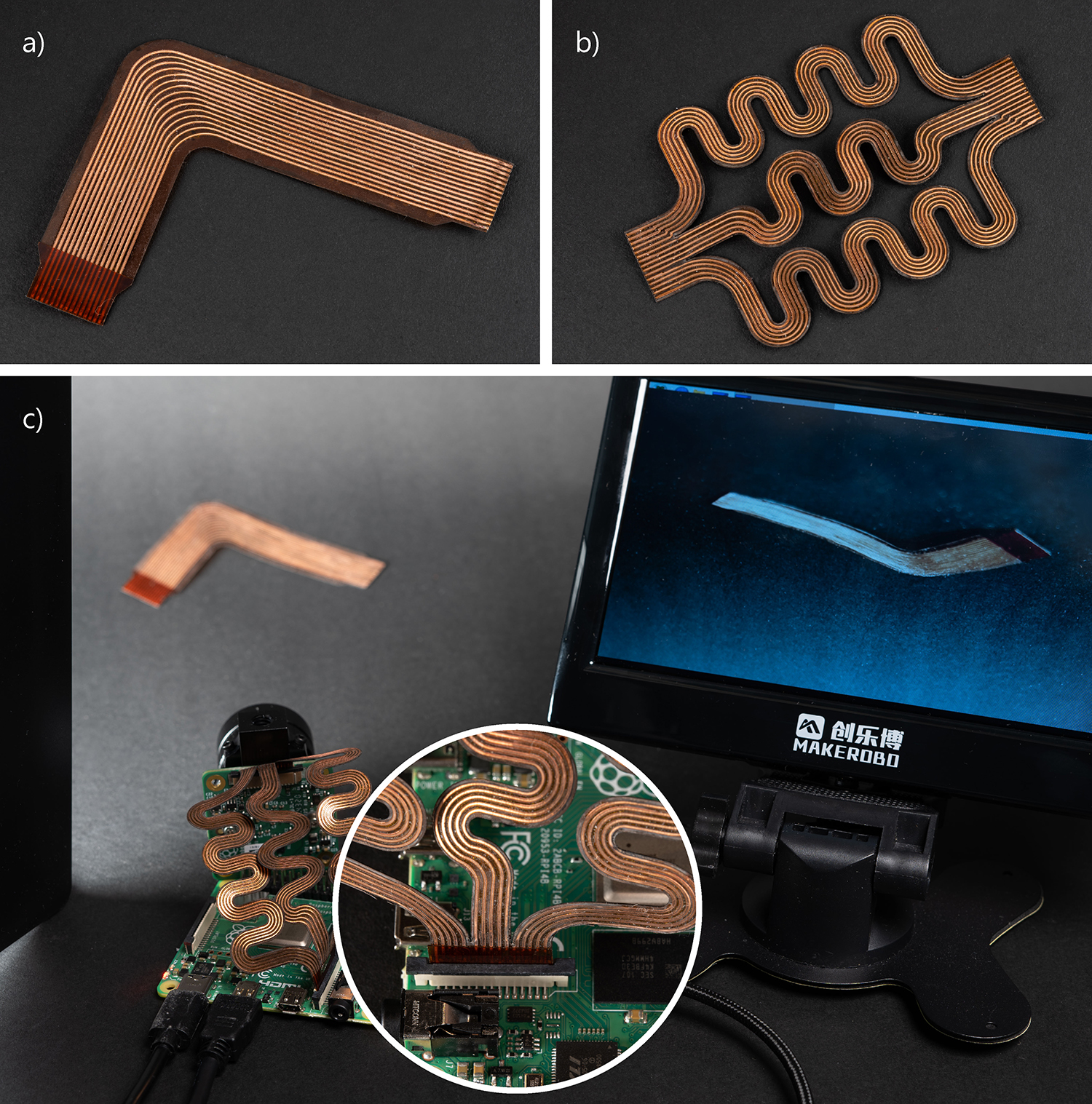}
\caption{Flex cables fabricated using Fibercuit. a) Right angle flex connector. b) Stretchable flex connector. c) Stretchable flex connector between a Raspberry Pi and a Pi camera module. }
\label{fig:flexcable}
\end{figure}

Due to the flexible nature of the circuits and the high resolution of Fibercuit, we can fabricate custom flex cables on demand. In this example, we demonstrate 2 cables --- a classic right-angle cable and a stretchable cable that is rare to find off-the-shelf (Figure \ref{fig:flexcable}a and b). Both cables are designed to work with a Raspberry Pi and a 15-pin Pi camera module (Figure \ref{fig:flexcable}c). 
The custom, stretchable cable grants flexibility for designers during the prototyping phase, as a designer can experiment with different camera placements with the stretchable cable set instead of being restricted to a pre-chosen length. 
 
\subsection{Custom End-Stop Switch} \label{endstop}

Fibercuit can be used to make custom sensors with integrated electrical and mechanical features in one package. Here, we demonstrate a small end-stop switch (footprint:  13 mm by 7 mm) that is laser-folded using Fibercuit (including laser forming the sensor). The two contact leaves are isolated by default, and the circuit is complete when a force is applied to push one of the copper leaves. An additional LED is added as part of the switch circuit to indicate its status. This mechanism can be extended to contact sensors and pressure sensors with custom form factors as required by the user. 

\begin{figure}[h]
\centering 
  \includegraphics[ width=\columnwidth]{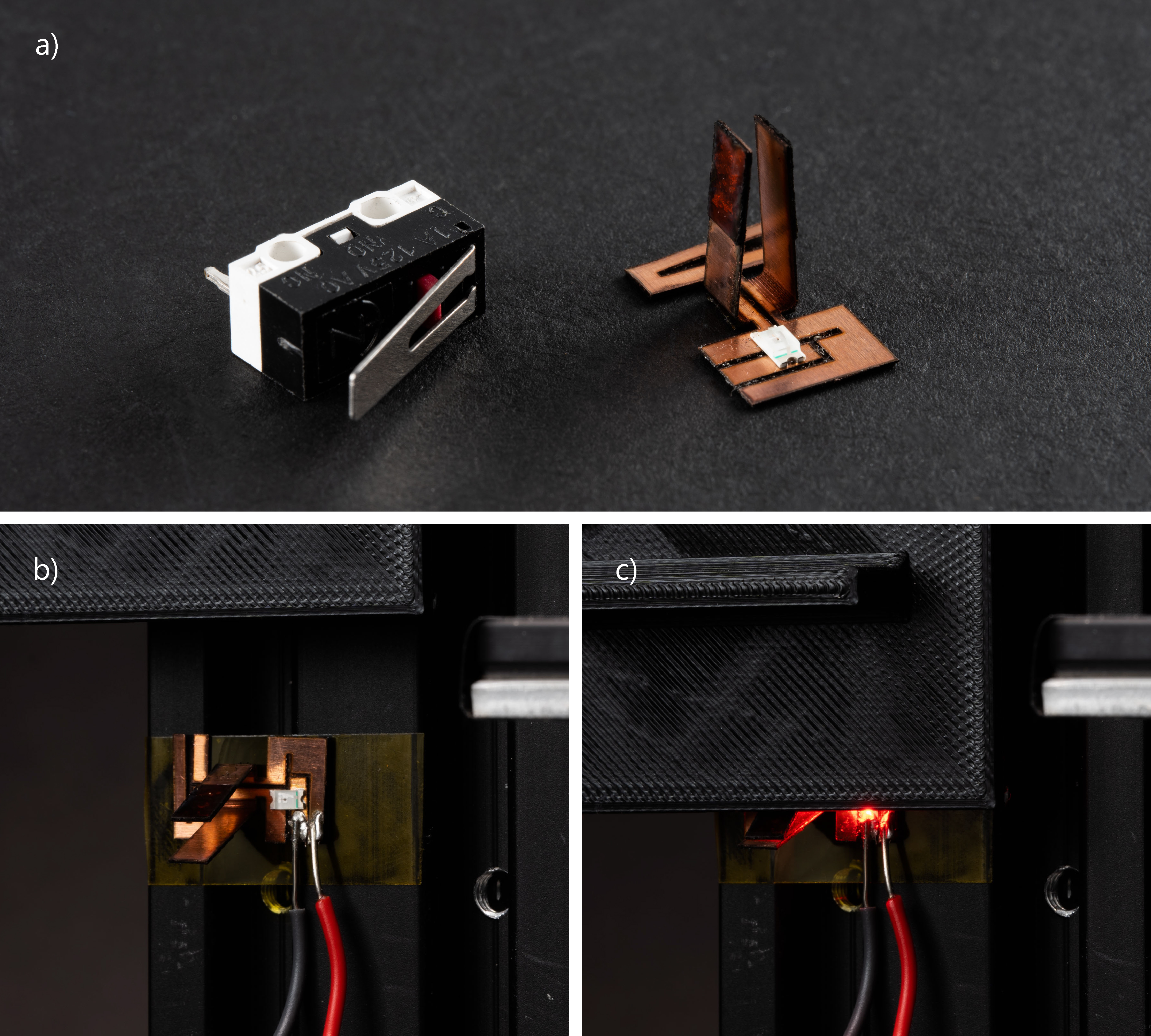}
\caption{a) A custom end-stop switch with integrated LED, fabricated using Fibercuit,  alongside a standard end-stop switch. b) End-stop switch installed on a 3D printer. No connection when the contact leaves are apart. c) The LED glows when the contact leaves touch each other.}
\label{fig:endstop}
\end{figure}

\subsection{Electromagnetic Coils}

\begin{figure}[h]
\centering 
  \includegraphics[ width=\columnwidth]{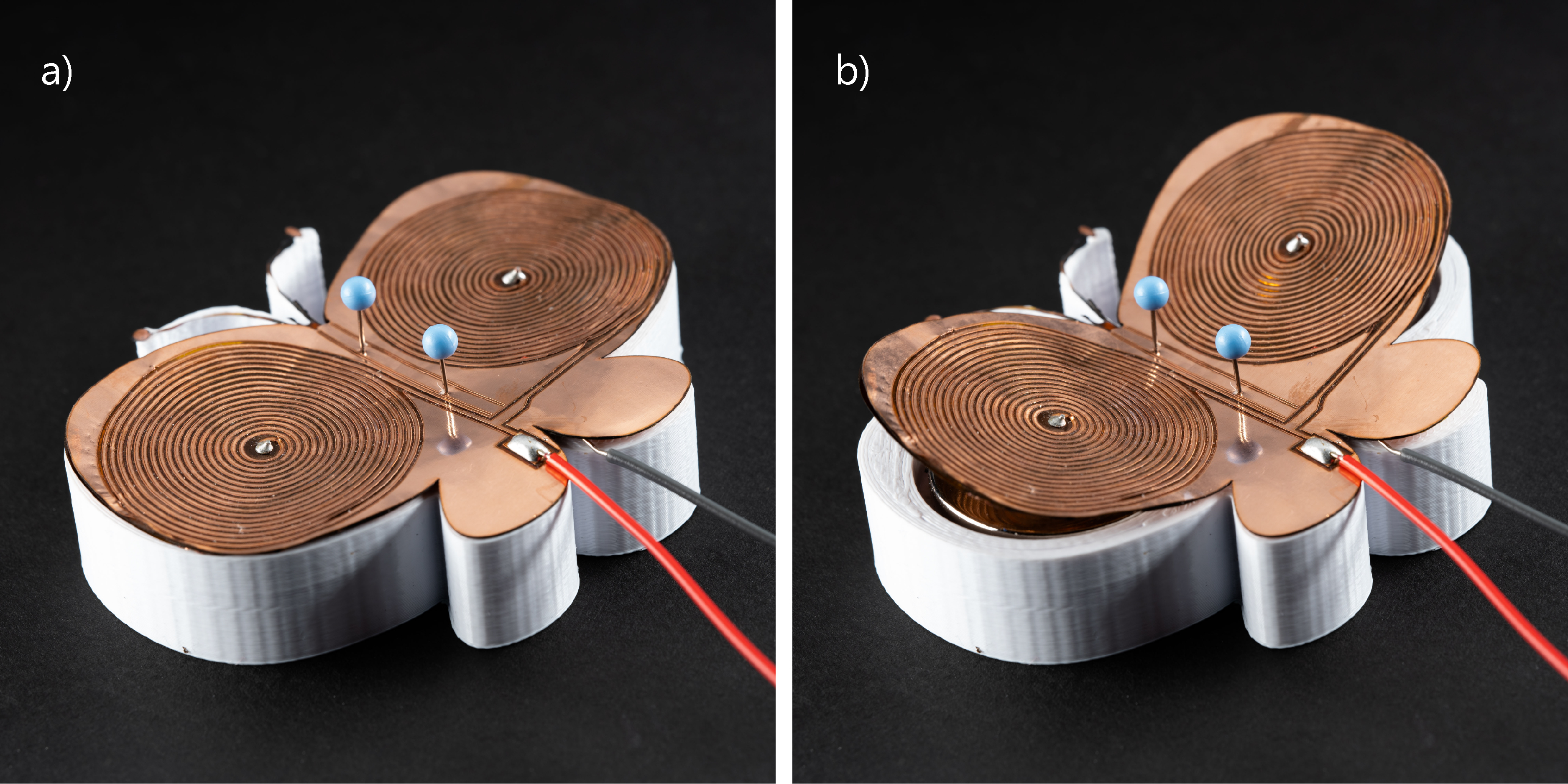}
\caption{A flapping actuator in the shape of a butterfly. a) Resting state. b) Repelling state. }
\label{fig:butterfly}
\end{figure}

An electromagnetic coil has multiple applications as actuators, inductance sensors, buzzers, and even prototype motors. Using Fibercuit, we can fabricate flexible 2-layer electromagnetic coils in custom form factors. We showcase this by creating a flapping actuator in the shape of a butterfly (Figure \ref{fig:butterfly}) where the electromagnetic wings repel the magnets in the 3D printed structure underneath to generate a flapping motion. This butterfly can also act as a speaker when connected to an audio amplifier. 

\subsection{LED Earrings}

\begin{figure}[h]
\centering 
  \includegraphics[ width=\columnwidth]{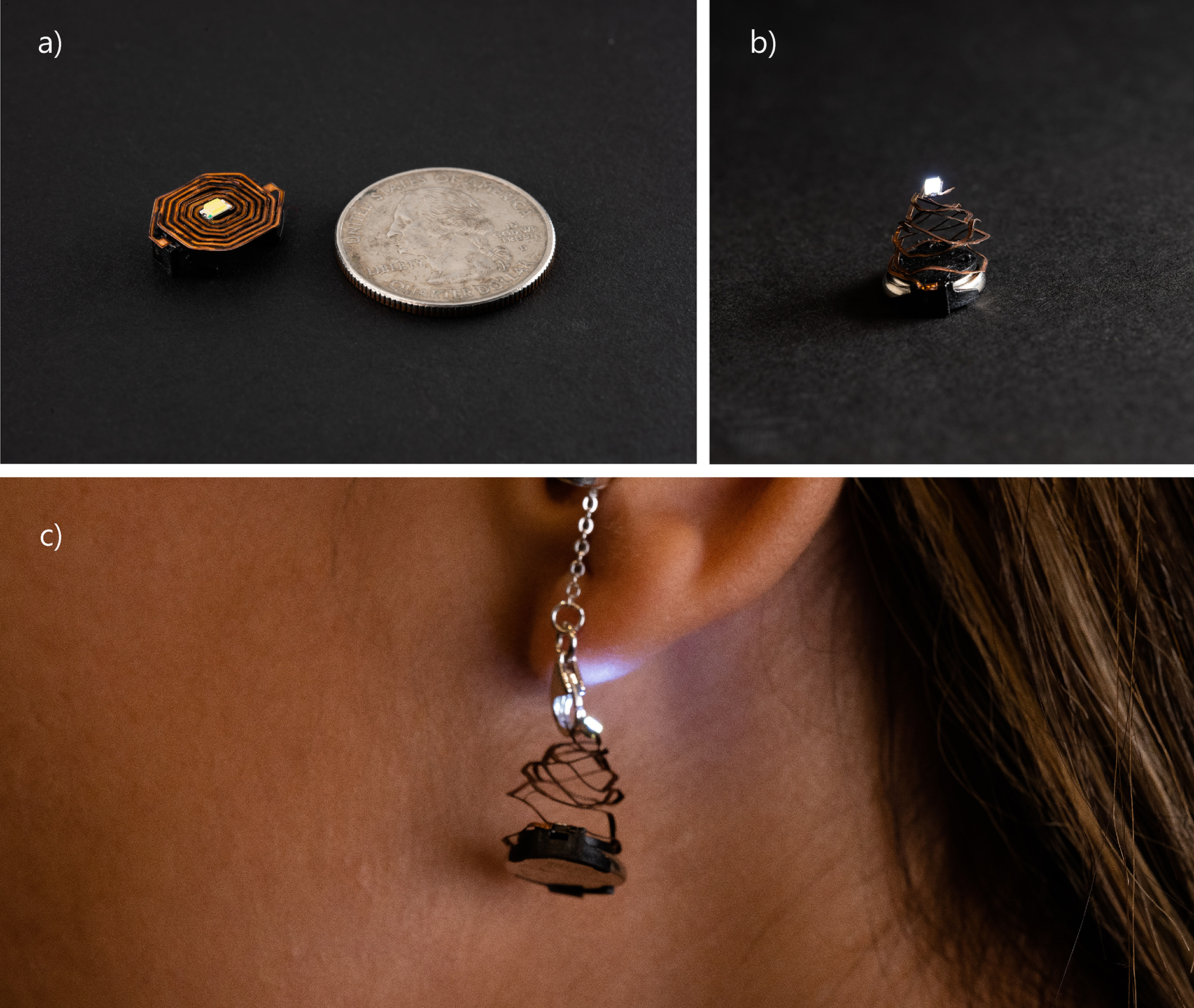}
\caption{a) Demonstrating the small size of the circuit using a quarter. b) Shape of the earring when the circuit is expanded. c) Showcasing the custom earring.}
\label{fig:earring}
\end{figure}

Figure~\ref{fig:earring} demonstrates a custom dome-shape earring with a size of 15 mm $\times$ 15 mm $\times$ 20 mm. A coin battery provides the weight required to deform the earring into the dome shape and is also used to light up the LED. This earring demonstrates how Fibercuit can enable the fabrication of miniature wearables by consolidating the embedded electronics and the form factor of the device. 

\subsection{Kirigami Crane} \label{crane}
\begin{figure}[h]
\centering 
  \includegraphics[width=\columnwidth]{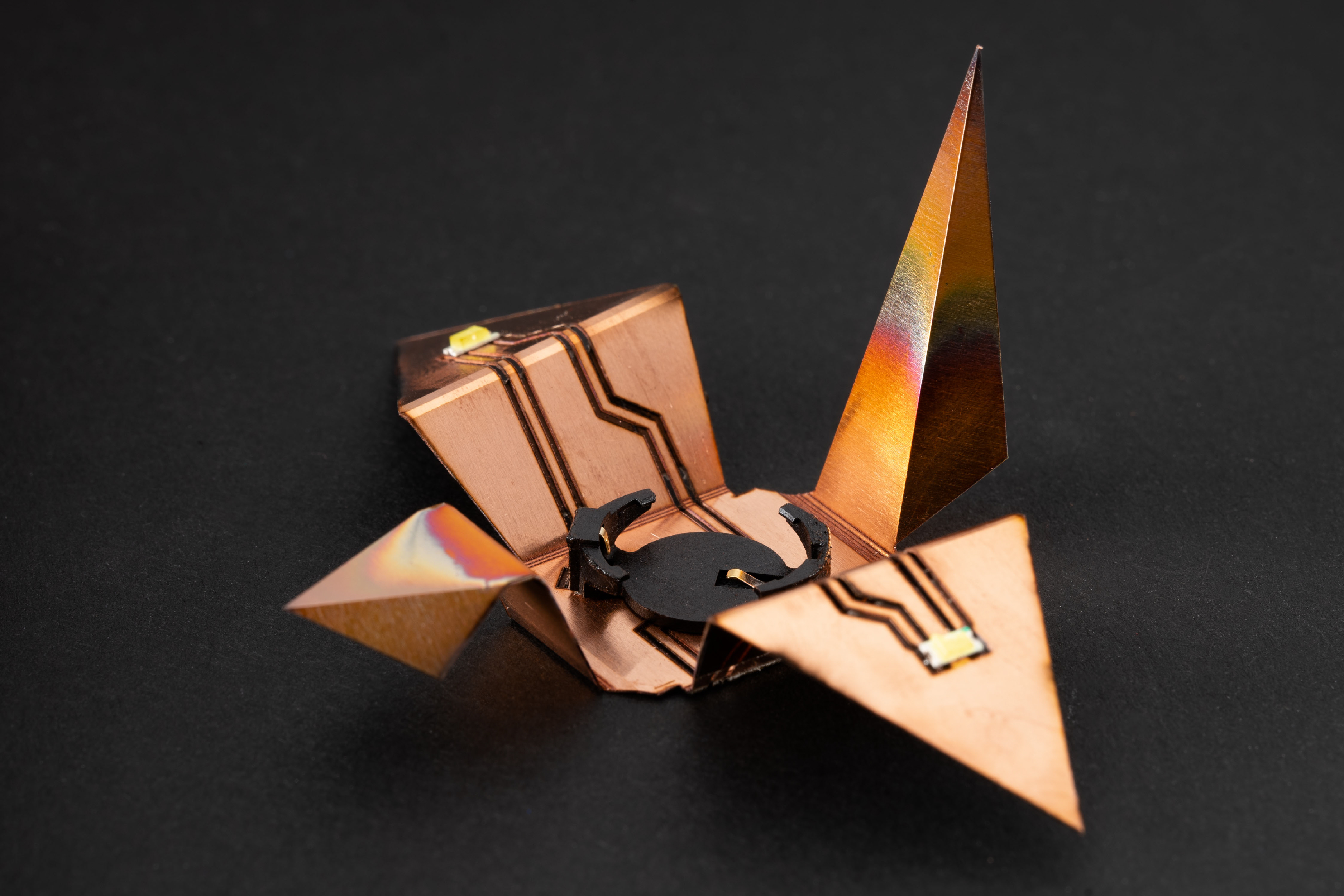}
\caption{A kirigami crane fabricated using Fibercuit. Contains a battery holder and 2 LEDs.}
\label{fig:crane}
\end{figure}

The addition of laser forming (Section \ref{laserforming}) enables us to explore kirigami circuits like the crane pictured in Figure \ref{fig:crane}. After fabricating the circuit and soldering the components onto it, the circuit is placed onto the bed and laser formed to form the kirigami crane with a battery holder and 2 LEDs. The ability to create kirigami structures could enable the prototyping of custom aesthetic wearables that appeal to users.

\section{User interface} \label{ui}
Most examples shown in the previous section can be designed using conventional circuit design tools such as EAGLE. However, since Fibercuit also supports laser forming kirigami circuits, we developed an alternative circuit design interface that incorporates laser forming simulation. This is to help end-users to have a preview of the folded circuits prior to fabrication. %

\subsection{Interface}

\begin{figure}[htb!]
\centering
  \includegraphics[width=\columnwidth]{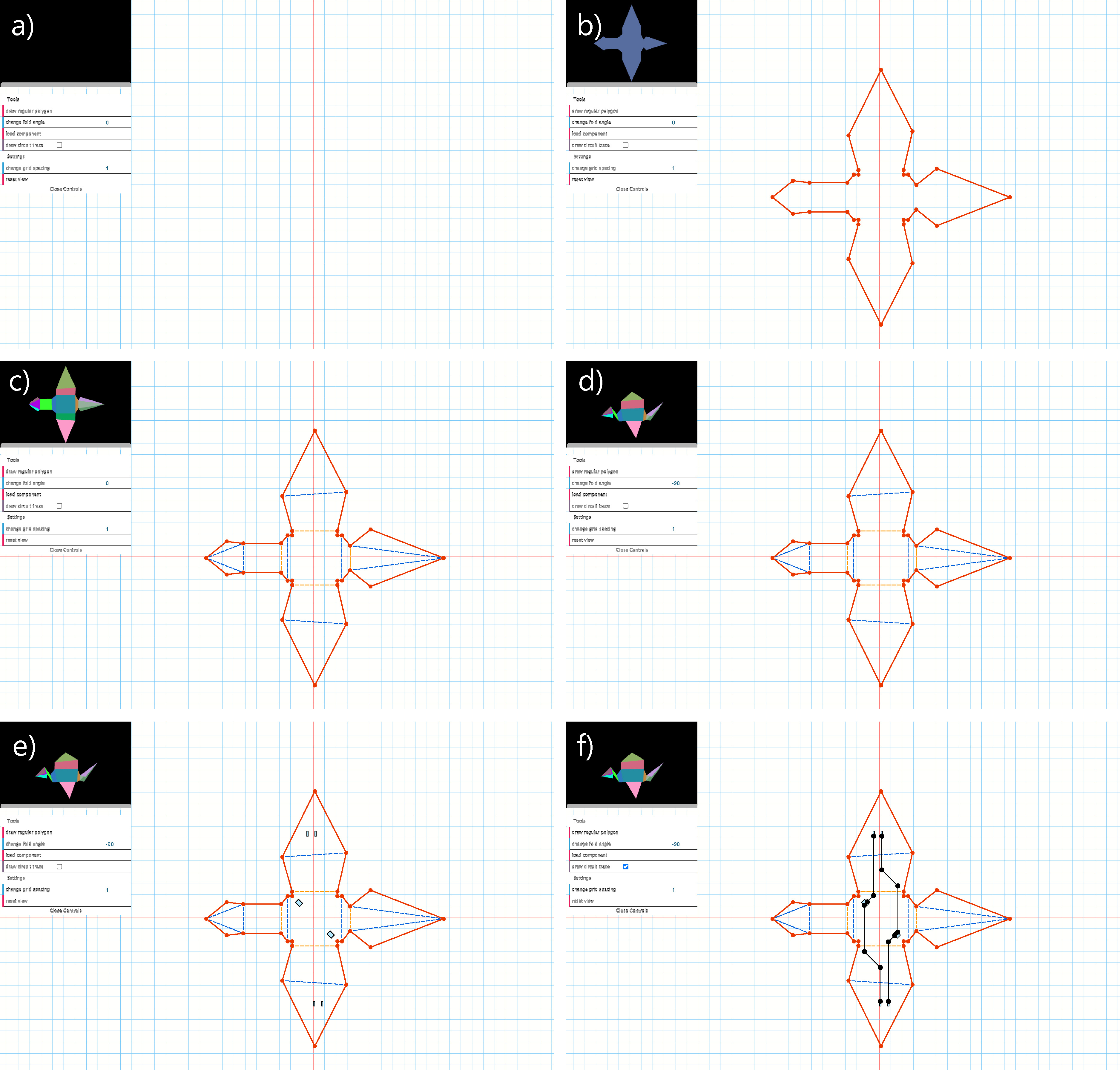}
  \caption{User interface to design and simulate the folding of a kirigami circuit.}~\label{fig:interface}
\end{figure}

The interface is written using the Three.js graphics library. Figure~\ref{fig:interface}a shows the grid-based 2D canvas upon which the user can draw three different types of edges: cutting edges, folding edges, and circuit trace edges. Folding edges can be further subdivided into mountain and valley edges, depending on the target angle at that location. The user interface visually reflects the angles at folding edges by increasing or decreasing the thickness of the corresponding lines, depending on the magnitude of the angle. 

In order to assist the user's understanding of how the laser forming will work, there is a 3D viewport included in the upper left corner of the interface which shows the shape of the folded design given the angles the user has set for each of the folding edges. This helps to detect designs which are not foldable by a laser cutter due to self-collisions or other issues.

The first step in creating a design, shown in figure ~\ref{fig:interface}b, is to create an outline by placing points along the border of the overall shape of the object. Cutting lines are automatically generated between these points to create a closed polygon. Figure~\ref{fig:interface}c displays the next step of creating the folding edges between certain points in the outline. This is accomplished by selecting points on the outline and using the keyboard shortcut to ``draw folding edge". To create a 3D, foldable shape, these edges can have their angles set to values between -90 and 90 degrees, depending on whether they are mountain or valley edges. This is shown in figure~\ref{fig:interface}d, along with the viewport which reflects folding progress at different states using the slider. Figure~\ref{fig:interface}e shows that once a circuit outline is created, the user can load electrical components represented as SVG footprints to be placed onto the canvas. In order to connect two pads, a user may create a new circuit trace line between them. The final design which includes these traces is shown in Figure~\ref{fig:interface}f.

\subsection{Cutting Profile Generation}

When the design is finalized, it can be exported as a multi-layer SVG file which contains all the relevant information ready for cutting. Lines are exported with related information such as edge type and folding angle, and are stored in different layers of the SVG. The layers are ordered the same as the cutting operations orders, based on the algorithm presented in~\cite{hao2021metal}. The SVG can be sent directly to the fiber laser for circuit fabrication.

\begin{table*}[h]
\caption{Laser Engraving Parameters}
  \label{tab:paras}
  \begin{tabular}{c  >{\centering\arraybackslash}p{0.19\textwidth} >{\centering\arraybackslash}p{0.05\textwidth} >{\centering\arraybackslash}p{0.05\textwidth} >{\centering\arraybackslash}p{0.04\textwidth} >{\centering\arraybackslash}p{0.12\textwidth} >{\centering\arraybackslash}p{0.16\textwidth}}
    \toprule[2pt]
    technique & sub-division & speed (mm/s)  & Power (\%) & pass & cooling time between pass (s) & profile\\
    \midrule[1.5pt]
    \multirow{5}{*}{cutting isolation outline} & 0.03mm copper & 500 & 60 & 14 & 5 & \multirow{5}{*}{outline of isolation area}\\

    \cline{2-6}
    
    &0.05mm copper & 500 & 60 & 25 & 10  & \\

    \cline{2-6}
    
    & 0.1mm copper & 200 & 60 & 14 & 10 & \\
    
    \cline{2-6}
    
    & 0.15mm copper & 100 & 60 & 15 & 15 & \\
    \cline{2-6}
    
    & 0.2mm copper & 60 & 60 & 28 & 15 & \\
    
    \midrule
    
    \multirow{2}{*}{cutting through} & \multirow{2}{*}{--} & \multirow{2}{*}{100} & \multirow{2}{*}{100} & \multirow{2}{*}{20} & \multirow{2}{*}{no cooling} & outline of hole, drill and dimension\\
    
    \midrule
    
    \multirow{2}{*}{solder mask removal}		& \multirow{2}{*}{--} & \multirow{2}{*}{500} & \multirow{2}{*}{30} & \multirow{2}{*}{3} & \multirow{2}{*}{no cooling} & SMD \& through-hole solder pad area raster\\
    
    \midrule
    
    \multirow{8}{*}{laser soldering}	 &through hole components and via	& \multirow{2}{*}{200} &	\multirow{2}{*}{100}	& \multirow{2}{*}{50} &	\multirow{2}{*}{no cooling} & 0.1mm - 0.3mm diameter circle\\
    
    \cline{2-7}
    
    & SMD component with extended solder pins &\multirow{2}{*}{500}		& \multirow{2}{*}{20} &	\multirow{2}{*}{50x3} &	\multirow{2}{*}{5s between sets}	& \multirow{2}{*}{0.3mm square raster}\\
    
    \cline{2-7}
    
    & SMD components without extended pin components &	\multirow{2}{*}{500}	& \multirow{2}{*}{30}	&	\multirow{2}{*}{100}&	\multirow{2}{*}{no cooling}&	0.3mm diameter hole raster \\

    \midrule
    
    \multirow{2}{*}{laser bending}  &	0.15mm copper &	100 &	45  & 30  & 0.5 & bending hinge\\
    
    \cline{2-7}
    
    & 0.2mm copper &	100 &	42  & 30  & 0.5 & bending hinge\\
    
    \midrule
    
    \multirow{2}{*}{Kapton removal} & \multirow{2}{*}{--}& \multirow{2}{*}{1000}	&\multirow{2}{*}{25}	&\multirow{2}{*}{30}	&\multirow{2}{*}{no cooling}	& Kapton under the bending hinge\\

    \bottomrule[2pt]
\end{tabular}
\end{table*}

\section{Machine and Process Parameters}

In this section we summarize the parameters of the machine and the modifications that were made to enable our fabrication workflow. 

\subsection{Machine and its Add-ons}

\begin{figure}[h!]
\centering
  \includegraphics[width=\columnwidth]{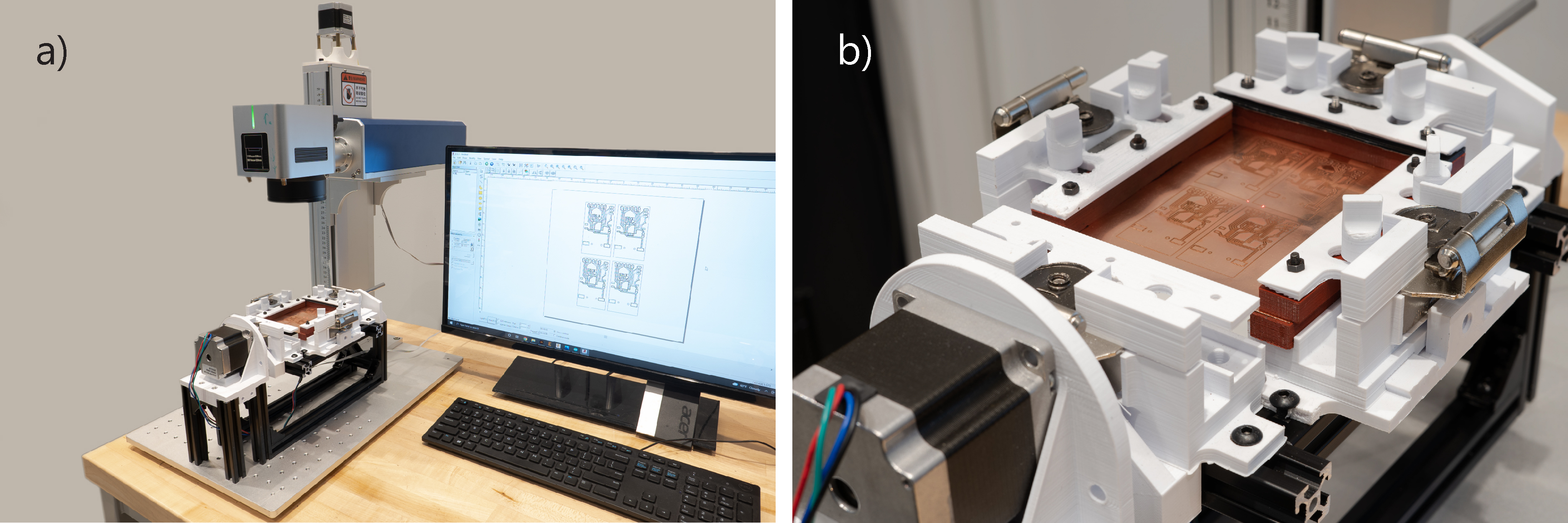}
  \caption{a) Modified 50W fiber laser engraver. b) Custom clamping platform.}~\label{fig:machine}
\end{figure}

Figure \ref{fig:machine}a shows the fiber laser engraver machine used in the paper.
Figure \ref{fig:machine} b shows the rotary clamping platform as a custom add-on.  
The clamp holds the composite material sheet down and stretches it in two orthogonal directions, ensuring the least amount of wrinkle and buckling.
3D printed dowel pins minimize the calibration error when the material has to be taken off the bed and put back in during the manufacturing process.
An integrated micro-controller is added to digitally control the rotation of the platform for the laser to perform dual-layer cutting and forming.  

\subsection{Process Parameters}

The process parameters mentioned in this paper are based on thorough experimentation on the 50W laser engraver. They include 1) percentage of power: percentage of power out of 50w used in engraving; 2) scanning speed: how fast the laser beam moves along the irradiation path and 3) cooling time between paths.
1) and 2) can be assigned to different profiles in the laser driver software with trace color visualization similar to CO2 laser drivers.
The cooling time is implemented using a dummy scan --- a 0\% power 10 mm long straight line segment passes with a fixed speed of 10 mm/s. 
For example, inserting 15 dummy scans between engraving passes generates 15s of cooling time.

The process parameters for major techniques are presented in table~\ref{tab:paras}, which includes parameters for cutting the conductor, laser forming, laser soldering, cutting through the board stock and solder mask removal.

\section{Discussion} \label{discussion}

\subsection{Effects of the Material Thickness} \label{thickness}

The thickness of the material influences the process parameters.
The required energy density is generally proportional to the thickness of the conductor being used.
This energy density can be increased by reducing the speed of the laser, increasing the power, and increasing the number of passes.
This change in the process parameters requires different considerations in the conducting layers and the dielectric layers. 

In order to accommodate laser forming, a thicker copper is preferred, as the temperature gradient mechanism performs better when the ratio between the material thickness and the beam diameter is higher. 
But when the energy density is increased, this leads to higher carbonization of the silicone adhesive on the underside of the Kapton. 

As a result, in our experiments, to achieve a satisfactory trade-off between the two, we relied on copper between 0.1mm and 0.15mm in thickness. 
This creates satisfactory insulation but also allows various bending angles during laser forming. 
Further testing on higher-powered fiber laser engravers may produce different outcomes.

\subsection{Copper Adhesion and Deterioration}

Unlike an industrial method, the copper layer in Fibercuit is not chemically adhered to the dielectric layer. 
As a result, over time, traces with isolated and miniature footprints have the tendency to peel away from the silicone adhesive. 
This effect can be seen in Figure \ref{fig:mask}. The small pads on the ATMega168's footprint that are not connected to any trace are peeled away (Section \ref{cutting}).
This issue is not a hindrance in a rapid prototyping workflow, as it does not affect the actual circuit functionalities.
However, to support long-term deployment, future research is needed on permanently affixing the traces to the dielectric substrate.

\subsection{N-layer Circuits}
The current via design in our procedure only affords the fabrication of dual-layer circuits. 
We note that some work in this area (\textit{e.g.}, \cite{multi_layer_flex_pcb}) has explored variations in via design, which could potentially 
extend the Fibercuit procedure for n-layer circuits fabrication. However, further explorations is needed to confirm the feasibility.

\subsection{Through-Hole Components}

In a standard PCB manufacturing process, the circuit board goes through an additional electrolytic copper plating procedure that deposits copper on plated through holes to interconnect between the conductive layers. 
As this step is missing in our procedure, through-hole components are soldered directly on the top layer of the circuit, rather than the bottom. 
Although this does not heavily affect a prototyping workflow, new designs on the via fabrication process, as discussed in the previous section, might enable the creation of interconnects in plated through holes. 

\subsection{Fabrcation Time}
As mentioned in section \ref{circuitSample}, the fabrication time is mostly confounded by human operations instead of machine time. 
In future work, we may explore automated techniques to reduce the manual operation, \textit{e.g.}, selectively removing the conductive layer material to substitute the manual peeling process, or adding a pick-and-place mechanism to the machine to automate the laser soldering process fully. 
With these in place, the fabrication time will be further reduced.

\section{Conclusion}
We presented Fibercuit, a novel set of fabrication techniques for making high-resolution, flexible, and kirigami circuits on-demand using a fiber laser engraver. We presented the composition and fabrication of custom copper-based composites and the laser cutting techniques to form fine-pitch conductive traces from them. We also discussed techniques that enable the fiber laser to solder small surface-mount electrical components and the ability to laser form kirigami circuits. Combined with our software pipeline, an end-user can design and fabricate flexible circuits that are dual-layer and kirigami circuits that can fit into a wide range of form factors.

\begin{acks}
We thank Singh Sandbox at the University of Maryland for the support on sample fabrication and video shooting. We also thank reviewers for their valuable feedback.
\end{acks}

\bibliographystyle{ACM-Reference-Format}
\bibliography{sample-base}

\end{document}